\documentclass[prd,11pt,nofootinbib]{revtex4}

\usepackage{amsmath}    
\usepackage{graphicx}   
\usepackage{subcaption} 
\usepackage{verbatim}   
\usepackage{color}      
\usepackage{hyperref}   
\usepackage{footnote}

\usepackage{adjustbox}  

\newcommand{\be}{\begin{equation}}
\newcommand{\ee}{\end{equation}}
\newcommand{\bea}{\begin{eqnarray}}
\newcommand{\eea}{\end{eqnarray}}
\newcommand{\nn}{\nonumber}

\begin{document}

\title{Constraining New Physics with Single Top production at LHC  }
\author{Daniel Stolarski}
\author{Alberto Tonero} 
\affiliation{Ottawa-Carleton, Institute for Physics, Carleton University
1125 Colonel By Drive, Ottawa, ON, K1S 5B6, Canada\\}

\begin{abstract}
We study effects of beyond the Standard Model physics coupling third generation quarks to leptons of the first two generations.
We parametrize these effects by dimension-six effective operators, and we also consider related simplified UV completions: scalar leptoquark and $W'$ models.
We derive new constraints on these
scenarios by using recent ATLAS measurements of differential cross sections of single top production in association with a $W$ boson, and also show how these limits will evolve with future data. 
We also describe how the limits can be significantly improved by using ratios of differential distributions with different flavours of leptons. \end{abstract}

\maketitle

\newpage

\tableofcontents
\newpage

\section{Introduction}
The LHC has now taken a significant amount of data at the TeV scale, but as yet it has found no evidence for physics beyond the Standard Model (BSM). While the searches and measurements performed at the LHC cover a tremendous range of BSM theories, it is not necessarily the case that the space of all possible observable deviations from the Standard Model (SM) has been searched for in the data. Furthermore, the LHC data set will grow significantly over the coming years, so it is critical to explore all possible BSM theories that could be discovered.

The top quark is a particularly interesting sector of the SM being the heaviest known fundamental particle and coupling most strongly to the electroweak symmetry breaking sector. Furthermore, the strong constraints from flavour physics on the other fermions of the SM are much weaker when applied to the top. And finally, the production cross section for the top at the LHC is quite large, so precision measurements can be made with present and future data with potentially groundbreaking sensitivity to BSM physics.

One way to classify BSM models where new particles are too heavy to be directly produced at LHC and therefore small deviations from the SM predictions are expected at low energy, is through the Standard Model Effective Field Theory~\cite{Weinberg:1979sa,Buchmuller:1985jz,Grzadkowski:2010es} (SMEFT). New physics effects are parametrized by a set of higher-dimensional operators organized in a series expansion, with increasing operator dimension.
Even at dimension 6, there are 2499 operators, so it is very difficult to make general statements. One strategy commonly taken, is to assume flavour universality and baryon number conservation, which reduces the number of operators to 59~\cite{Grzadkowski:2010es}. This strategy, however, does not allow one to study physics coupled predominantly to the top quark.

The classification of effective operators contributing to top quark processes~\cite{AguilarSaavedra:2008zc,Zhang:2010dr,AguilarSaavedra:2010zi} marked the beginning of a significant research activity devoted to find novel ways to constrain higher dimensional operators involving top quarks using LHC data. The processes that have been considered include top pair production~\cite{AguilarSaavedra:2010zi,Aguilar-Saavedra:2014iga,Schulze:2016qas,Barducci:2017ddn,Martini:2019lsi}, top decay~\cite{Durieux:2014xla,Chala:2018agk}, and top production in association with a Higgs~\cite{Maltoni:2016yxb,Degrande:2018fog} a $Z$ or $\gamma$~\cite{Schulze:2016qas,Degrande:2018fog,Tonero:2014jea}, or $b$ quarks~\cite{DHondt:2018cww}. There has also been a study using low energy observables~\cite{Cirigliano:2016nyn} as well as a future lepton collider~\cite{Durieux:2018ggn}. In addition, there are two groups, TopFitter~\cite{Buckley:2015nca,Buckley:2015lku,Brown:2019pzx} and SMEFiT~\cite{Hartland:2019bjb,Brivio:2019ius} that have performed global fits to data to constrain the space of EFT operators coupling to tops. Finally, there is a review~\cite{AguilarSaavedra:2018nen}, that compiled the latest constraints on the complete set of operators involving top quarks (modulo some assumptions about flavour). 

In this work we consider the process of top quark production in association with a $W$ boson as an avenue to constrain as yet unconstrained operators in the SMEFT. This can be done thanks to the ATLAS measurement~\cite{Aaboud:2017qyi} of \textit{unfolded differential} cross sections of $pp\to tW \to b l^+\nu l^-\bar \nu$, which was not used in the global fits of~\cite{Buckley:2015nca,Buckley:2015lku,Brown:2019pzx,Hartland:2019bjb,Brivio:2019ius}. The fact that the measurement is unfolded means that the experimental uncertainties are removed from the final results, allowing us to compare it to theoretical calculations of the cross sections at particle level. The differential nature of the measurements is also crucial, because new physics, especially when parameterized via effective field theory (EFT), will mainly show up in the high energy tails of distributions, while the SM contributions will be largest in the phase space closer to the production threshold. Therefore, differential measurements of the sorts in~\cite{Aaboud:2017qyi} can place novel constraints. 

In practice, measurements such as~\cite{Aaboud:2017qyi} are sensitive to the full $tW$ final state that consists of leptonic (electron or muon) decays of the $W$ and top quark because the $W$ or the top are not reconstructed. Therefore, this analysis searches for a final state consisting of 
\begin{itemize}
\item One $b$-tagged jet
\item Exactly two leptons of opposite charge
\item Missing energy
\item No additional hard jets ($b$-tagged or not),
\end{itemize}
which will be sensitive the process $pp \rightarrow tW \rightarrow b \ell^+ \ell^- \nu\bar{\nu}$, where $\ell=e,\mu$. In this study, we will therefore focus on new physics that can contribute to this process, and thus couples to top quarks as well as first or second generation leptons. We will ignore new physics coupling to $\tau$, but see~\cite{Kamenik:2018nxv} for a recent work. 
The impact of contact interactions involving two leptons (electrons or muons) and two $b$-quarks in di-lepton final state was also recently studied in~\cite{Afik:2019htr}.
	
In this work, we will show that current data places a constraint on the scale $\Lambda$  suppressing the new physics operators which reads $\Lambda \gtrsim $ few $\times\, 100$ GeV. Given the center of mass energy of the LHC, this bound is strictly outside the range of validity of the EFT. Therefore, we also consider and constrain (by means of the same measurements) two simplified UV completions of our effective operators. 

The first simplified model belongs to family of leptoquark (LQ) models~\cite{Dorsner:2016wpm} and assumes the presence of a single scalar field $R$ with SM gauge quantum numbers $(3,2,7/6)$ that couples only to third generation quarks and leptons of either the first or second generation. If the LQ had significant couplings to both the electron and muon, it would be excluded by strong constraints from $\mu\rightarrow e\gamma$~\cite{TheMEG:2016wtm}. 

Recent experimental constraints on pair production of scalar leptoquarks at LHC can be found in~\cite{Sirunyan:2018ryt,Sirunyan:2018ruf,Sirunyan:2018btu,Aaboud:2019jcc,Aaboud:2019bye}. Most searches assume a leptoquark that couples the $i$th generation of quarks to the $i$th generation of leptons ($i=1,2$, or 3) and do not apply to the scenario we are considering here. One exception is~\cite{Sirunyan:2018ruf}, which places limits on LQ coupling top quarks to muons of order 1.2 TeV (see also~\cite{CMS:2018yke} which studies projections for HL-LHC). The analysis of~\cite{Sirunyan:2018ruf} assumes that the LQ has electric charge $-1/3$ (as opposed to the model we consider with charge $5/3$), and this charge assumption is used to distinguish leptons that come from the decay of the LQ vs.~those that come from the decay of the top. Therefore, we expect the actual limit to be somewhat weaker, but a full recast is beyond the scope of this work.

Recent theory work on scalar leptoquark searches can be found in~\cite{Dorsner:2014axa,
Mandal:2015vfa,Diaz:2017lit,Bansal:2018eha,Monteux:2018ufc,Alves:2018krf,Dekens:2018bci,Schmaltz:2018nls,Chandak:2019iwj,Borschensky:2020hot}, with~\cite{Chandak:2019iwj} specifically focusing on leptoquarks that decay to top quarks and light charged leptons. The work of~\cite{Diaz:2017lit} uses a recast of the CMS multi-lepton search~\cite{CMS:2017iir} to set a limit of 800 GeV on this scenario. This scenario can also be constrained by LEP data~\cite{Arnan:2019olv} with the one-loop correction to $Z$ couplings to leptons providing a particularly strong constraint. 
In this work, for completeness, we will consider masses below these bounds, but the most interesting region is of course the region that is not excluded. 

The second simplified model we consider is a generalized sequential $W'$ model~\cite{Hsieh:2010zr} with non-universal couplings in flavour space where the charged gauge boson $W'$ couples only to the third generation left-handed quarks and first two generation left-handed leptons. Such sequential $W'$ models have been recently studied in the context of anomalies in $B$-physics~\cite{Greljo:2015mma,Boucenna:2016wpr,Boucenna:2016qad,Wang:2018upw,Zuo:2018sji}. Direct searches for such states typically look at couplings to only third generation quarks and leptons, so there are no such bounds these types of vectors that couple to third generation quarks and first or second generation leptons. Like the LQ model, if the $W'$ has generic couplings to muons and electrons, it will be excluded by the strong bounds on $\mu\rightarrow e\gamma$~\cite{TheMEG:2016wtm}. If, however, the couplings to all flavours of leptons are universal, there will be a GIM-like suppression of $\mu\rightarrow e\gamma$ just like for the SM $W$~\cite{Bilenky:1977du}. The $W'$ also enters at one loop and modifies the $Z$ boson decay, but how exactly it contributes depends on the specific symmetry breaking mechanism responsible for the $W'$ mass, which we have not specified.

In this work we place bounds on these models
and also estimate how the bounds will evolve with more LHC data. For concreteness, we assume the LQ or $W'$ only couples to electrons, but if the new state only couples to muons the bounds will be very similar because our analysis is approximately symmetric between $e$ and $\mu$. In the $W'$ case with universal couplings to all flavours, the bounds on the mass on the new physics will be approximately $\sqrt{2}$ stronger. 

If there is new physics of this type, UV considerations as well as strong bounds from $\mu\rightarrow e\gamma$ indicate that it generically couples dominantly to a single flavour of lepton. Therefore, inspired by~\cite{Greljo:2017vvb} and~\cite{Kamenik:2018nxv}, we also consider ratios of measurements in the electron vs.~muon channel. These ratios have partial cancellation of systematic errors and place significantly stronger constraints if new physics couples dominantly to one flavour of lepton. We explore the bounds these sorts of measurements could place, but as yet no such measurements exist in the literature. 

The organization of this paper is as follows. In Sec.~\ref{sec:EFT} we enumerate the EFT operators that contribute $tW$ production and explain the three we focus on in this work. In Sec.~\ref{sec:models} we describe two simplified UV completions that can be mapped onto our operators of interest and that we will also explore in this work. In Sec.~\ref{sec:sim} we outline our simulation framework, in particular how we compare to the results of~\cite{Aaboud:2017qyi}, and in Sec.~\ref{sec:results} we give the results achieved with current and future measurements for the EFT and for the simplified UV completions. In Sec.~\ref{sec:ratios}, we explore the improvements that can be obtained with new measurements using the ratios of distributions for different flavours, and we conclude in Sec~\ref{sec:conc}. Additional technical details are given in Appendix~\ref{app:fitting}.

\section{EFT }
\label{sec:EFT}
The language of Standard Model Effective Field Theory (SMEFT) is very suitable for phenomenological studies in presence of heavy BSM physics. In particular, when new physics degrees of freedom are much heavier than the energy scales relevant to single top production, one can describe the most general departures from the SM predictions in terms of higher dimension effective operators. The leading contributions come from dimension six operators~\cite{Grzadkowski:2010es} and can be parametrized in terms of an effective lagrangian as follows
\be 
{\cal L}_{\rm EFT}={\cal L}_{\rm SM}+\sum_i \frac{c_i}{\Lambda^2}{\cal O}_{i}
\ee
where $\Lambda$ represents the scale of BSM particles and $c_i$ are dimensionless Wilson coefficients.

Single top production at LHC in association with a lepton pair  $p p \to t l\bar \nu$ can be modified by the presence of these higher dimensional operators. Among all possible dimension six terms (59 modulo flavour) that  belong to the SMEFT lagrangian~\cite{Grzadkowski:2010es}, one can identify 8 operators (modulo flavour) that give rise to top-quark interactions that contribute at tree level to the process $p p \to t l\bar \nu$ (see~\cite{AguilarSaavedra:2018nen} for more details). We adopt the same notation and the same flavour assumption for the fermion bilinears as in~\cite{AguilarSaavedra:2018nen}:  flavour diagonality in the lepton sector and $U(2)_q\times U(2)_u \times U(2)_d$ flavour symmetry in the quark sector. Assuming this flavour symmetry, we can write down the relevant operators for $p p \to t l\bar \nu$ by grouping them into three different classes depending on the nature of the induced top-quark interactions. In class I, we have the following operators:
\bea \label{eftopclI}
\frac{c^3_{\varphi Q}}{\Lambda^2}\,(\varphi^\dagger i \overleftrightarrow D_\mu^I \varphi)(\bar Q_L\gamma^\mu \sigma^I Q_L)&&\qquad \frac{c_{\varphi tb}}{\Lambda^2}\,(\tilde \varphi^\dagger i  D_\mu \varphi)(\bar t_R\gamma^\mu b_R)+{\rm h.c.}\nn\\  \frac{c_{tW}}{\Lambda^2}\,\bar Q_L \sigma_{\mu\nu}\sigma^I t_R\tilde \varphi W_I^{\mu\nu} +{\rm h.c.}&&\qquad \frac{c_{bW}}{\Lambda^2}\, \bar Q_L \sigma_{\mu\nu}\sigma^I b_R \varphi W_I^{\mu\nu}+{\rm h.c.} \; ,
\eea
where $\varphi$ is the Higgs doubet, $Q=(t\quad b)$, and $W_{\mu\nu}^I$ is the $W$ field strength tensor. These operators induce anomalous $Wtb$ couplings that are parametrized in the literature by the following effective lagrangian \cite{AguilarSaavedra:2008zc}
\be 
\label{wtbv}
{\cal L}_{Wtb}=-\frac{g}{\sqrt{2}}\bar b\gamma^\mu (V_L P_L+ V_R P_R) t W_\mu^-
-\frac{g}{\sqrt{2}}\bar b \frac{i\sigma^{\mu\nu}q_\nu}{m_W}
(g_L P_L+ g_R P_R)t W_\mu^-+{\rm h.c.} 
\ee
The explicit contributions of the operators in Eq.~\eqref{eftopclI} to the anomalous $Wtb$ couplings are given by the following relations
\bea 
V_L&=&V_{tb}+c^{(3)}_{\varphi Q}\frac{\upsilon^2}{\Lambda^2} \qquad
V_R=\frac{1}{2}c_{\varphi tb}^*\frac{\upsilon^2}{\Lambda^2}\qquad
g_R=\sqrt{2}\,c_{tW}\frac{\upsilon^2}{\Lambda^2}\qquad
g_L=\sqrt{2}\,c_{bW}^*\frac{\upsilon^2}{\Lambda^2} \, .
\eea 
The operators in Eq.~\eqref{eftopclI} have been constrained in global fits using the full set of Tevatron and LHC Run I data that include total cross-sections as well as differential distributions, for both single top and pair production~\cite{Buckley:2015lku}. Comparable limits have been obtained also in studies that considered just anomalous $Wtb$ couplings at LHC, for an early study see~\cite{AguilarSaavedra:2011ct} while for more recent ones see~\cite{Fabbrichesi:2014wva,Cao:2015doa,Jueid:2018wnj}. Assuming an ${\cal O}(1)$ coefficient, the scale of new physics $\Lambda$ probed in these analyses varies from 400 GeV to 1 TeV, depending on the operator. 
  
In class II we have the top chromomagnetic dipole operator
\be \label{eftopsclII}
\frac{c_{tG}}{\Lambda^2}\bar Q_L \sigma_{\mu\nu}T^a t_R\tilde \varphi G_a^{\mu\nu}+{\rm h.c.}
\ee 
which is responsible for anomalous couplings to the gluon~\cite{Zhang:2010dr}. Stringent bounds on this effective operator coefficient can be found in more dedicated studies~\cite{Barducci:2017ddn,Aguilar-Saavedra:2018ggp}. Assuming an ${\cal O}(1)$ coefficient, the scale of new physics probed in these analyses is $\Lambda \gtrsim 1$ TeV. Finally, in class III we have the following four-fermion operators
\be\label{eftops}
\frac{{c}^{3(1)}_{Ql_i}}{\Lambda^2}(\bar l_{iL} \gamma^\mu \sigma^I l_{iL})(\bar Q_L \gamma_\mu \sigma_I Q_L)\qquad\frac{{c}^{S(1)}_{tl_i}}{\Lambda^2}(\bar l_{iL} e_{iR})\epsilon (\bar Q_L t_R)+{\rm h.c.}\qquad  \frac{{c}^{T(1)}_{tl_i}}{\Lambda^2}(\bar l_{iL} \sigma^{\mu\nu}e_{iR})\epsilon (\bar Q_L\sigma_{\mu\nu} t_R)+{\rm h.c.}
\ee
where $l_i=(\nu_i\quad e_i)$ and $i=1,2,3$ represents the lepton family index. 
While operators belonging to class I and II are already constrained by LHC + Tevatron measurements, class III operators of Eq.~\eqref{eftops} turn out to be currently unconstrained. In this work we want to fill this gap and we will use the recent ATLAS measurement of differential cross-sections of single top quark produced in association with a $W$ boson~\cite{Aaboud:2017qyi}, with 36.1 fb$^{-1}$, to put for the first time constraints on these effective operator coefficients. 

For concreteness, we assume that interactions involving the first generation leptons and third generation quarks. Our analysis is approximately symmetric between electrons and muons, so the limits in the scenario that couples dominantly to 2nd generation leptons instead would have very similar limits. If the new physics couples to both $e$ and $\mu$, it will be excluded by $\mu\rightarrow e\gamma$~\cite{TheMEG:2016wtm}, except for very specific flavour structures.

\section{Simplified models}
\label{sec:models}
The EFT description discussed above is only valid up to the mass scale of new physics $\Lambda$ where it should be matched onto a dynamical model involving new degrees of freedom.  If the higher dimension
operators are generated at tree level, then the matching implies the presence of new charged  particles. In this work we consider, in addition to EFT, two simplified UV models that induce modifications to the single top production $p p \to t l\bar \nu$ and can be matched into the operators of Eq.~\eqref{eftops}. These models are a scalar leptoquark (LQ) model and a $W'$ model, and will be presented in more detail here. We will use the recent ATLAS measurement of single top differential cross-sections~\cite{Aaboud:2017qyi} to put constraints in the {\it mass vs coupling} plane of these UV models.

\subsection{Scalar leptoquark model}
We consider the SM extended with a single scalar leptoquark $R$ of mass $M_R$ in the SM gauge group representation $(3,2,7/6)$. This is the only scalar leptoquark model that i) does not induce proton decay at tree level~\cite{Arnold:2012sd} and ii)  generates at low energy the EFT operators we are interested in. The most general Yukawa couplings to SM fermions can be parametrized as follows~\cite{Dorsner:2014axa}
\be \label{lagrlq}
{\cal L}_Y=z^*_{ij}\bar e_R^i R^{a*}Q_L^{j,a}-y^*_{ij}\bar u_R^i R^{a}\epsilon_{ab}l_L^{j,b}+{\rm h.c.}
\ee
where $i,j=1,2,3$ run over the fermion generations. To map this model onto the EFT flavour structure described in Sec.~\ref{sec:EFT}, we take only the $z_{13}$ and $y_{31}$ couplings to be real and non-vanishing. Furthermore, we assume $|z_{13}|=|y_{31}|\equiv g$. Therefore, integrating out the heavy leptoquark $R$ at tree level induces the following effective four-fermion operator
\be 
\pm\frac{g^2}{M_R^2}(\bar  l_L t_R)\epsilon(\bar Q_Le_R)+{\rm h.c.}
\ee
where the isospin and color contractions are not explicitly shown. Using the following Fierz relation for anticommuting fields
\be 
(\bar  l_L t_R)\epsilon(\bar Q_Le_R)=-\frac{1}{2}(\bar l_L e_R)\epsilon (\bar Q_L t_R)-\frac{1}{8}(\bar l_L \sigma^{\mu\nu}e_R)\epsilon (\bar Q_L\sigma_{\mu\nu} t_R)
\ee
we can see that we can match this model into the EFT operators of Eq.~\eqref{eftops} if 
\be \label{case1}
\frac{{c}^{3(1)}_{Ql}}{\Lambda^2}=0\qquad {\rm and} \qquad \frac{{c}^{S(1)}_{tl}}{\Lambda^2}=4\frac{{c}^{T(1)}_{tl}}{\Lambda^2}\equiv 4C_1
\ee
where
\be \label{match1}
C_1=\pm\frac{g^2}{8M_R^2} .
\ee
The minus (plus) sign corresponds to the case of same (opposite) sign $z_{13}$ and $y_{31}$ couplings.
While other UV completions are possible for the scalar and tensor operators of Eq.~\eqref{eftops}, we take this one to be representative.

\subsection{$W'$ model}
We consider the SM extended with an additional gauge boson $W'$ that couples only to left-handed leptons and quarks as follows
\bea \label{lagrwp}
{\cal L}&=&\frac{g_W}{\sqrt{2}}k_L^l\bar \nu_{Li} \gamma^\mu C_{ij}^{lL}e_{Lj}W'_\mu +\frac{g_W}{\sqrt{2}}k_L^q\bar u_{Li} \gamma^\mu  C_{ij}^{qL}d_{Lj}W'_\mu +{\rm h.c.}
\eea
where $g_W$ is the weak $SU(2)_L$ coupling, $k_L^l$ and $k_L^q$ are real rescaling factors and $i,j=1,2,3$ run over the fermion generations. Analogous to the LQ case, we assume that only the $C_{11}^{lL}$ and $C_{33}^{qL}$ couplings are non-vanishing and to further reduce the free parameters we assume $C_{11}^{lL}=C_{33}^{qL}=1$. In addition, we take  $|k_L^l|=|k_L^q|=k_L$. Therefore, integrating out the heavy $W'$ boson at tree level induces the following effective four-fermion operator
\be 
\pm\frac{g_W^2 k_L^2}{2 M_{W'}^2}(\bar \nu_L\gamma^\mu e_L)(\bar b_L\gamma^\mu t_L) +{\rm h.c.}
\ee
that can be matched into the EFT operators of Eq.~\eqref{eftops} if 
\be \label{case2}
\frac{{c}^{3(1)}_{Ql}}{\Lambda^2}\equiv C_2\qquad {\rm and} \qquad\frac{{c}^{S(1)}_{tl}}{\Lambda^2}=\frac{{c}^{T(1)}_{tl}}{\Lambda^2}=0
\ee
where
\be\label{match2}
C_2=\pm\frac{g_W^2 k_L^2}{4 M_{W'}^2}.
\ee
The minus (plus) sign corresponds to the case of same (opposite) sign $k_L^l$ and $k_L^q$ couplings.
We see that the EFT coefficient has the same parametric scaling here as in the LQ case: couplings squared divided by mass squared.

\section{Simulations}
\label{sec:sim}
\begin{figure}[h!]
\vspace{0.5cm}
\includegraphics[scale=0.25]{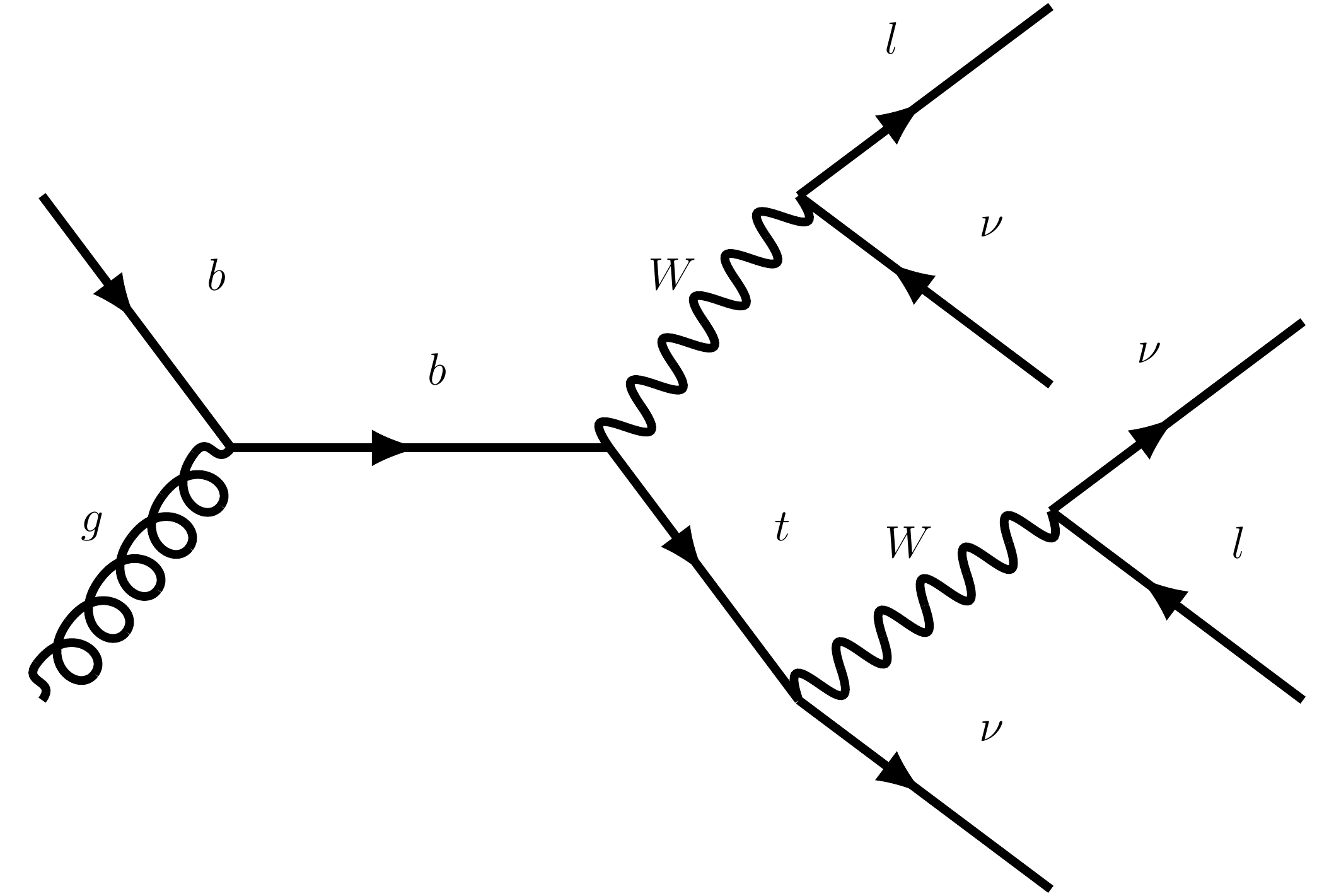} 
\includegraphics[scale=0.25]{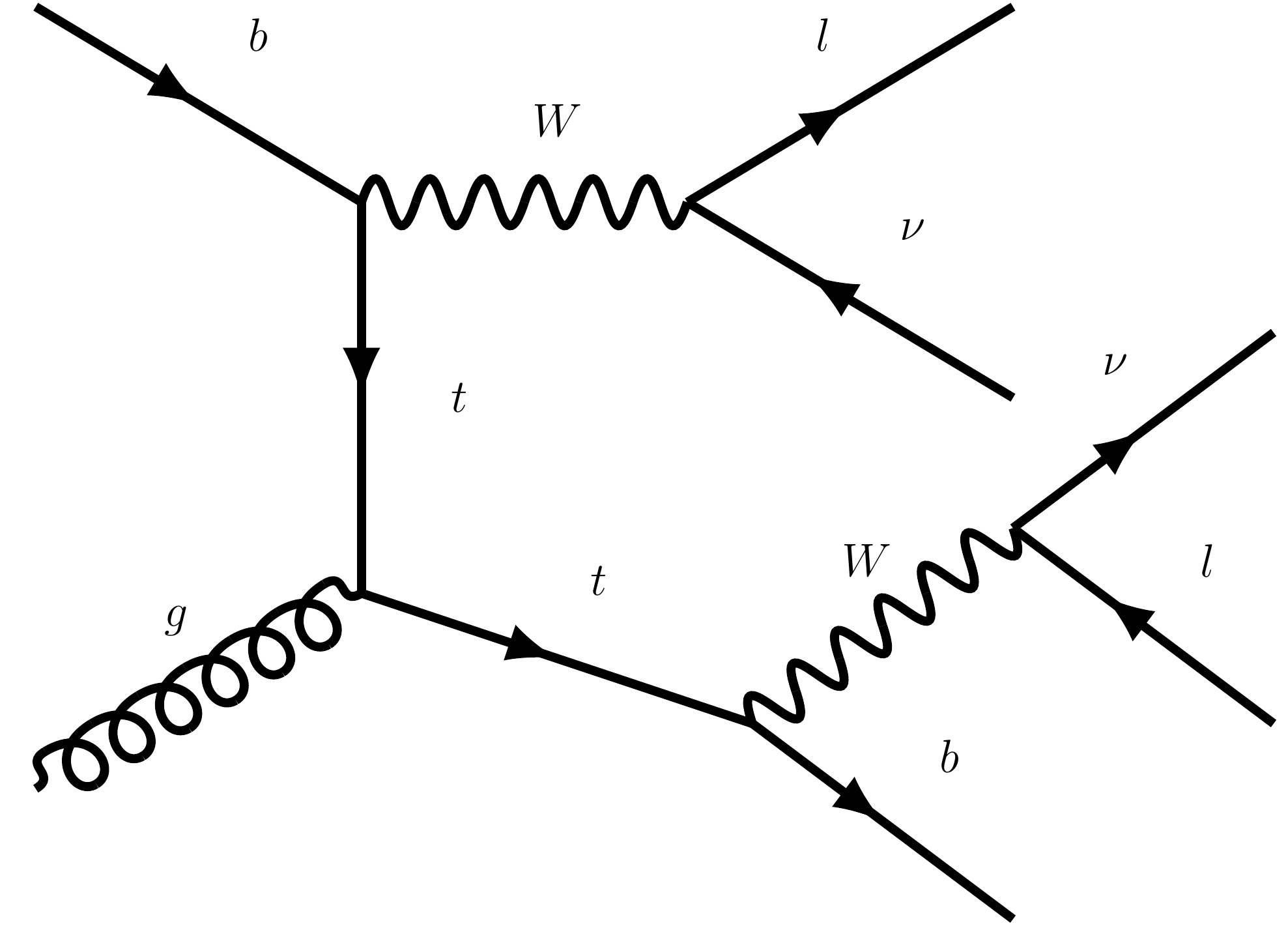} 
\includegraphics[scale=0.25]{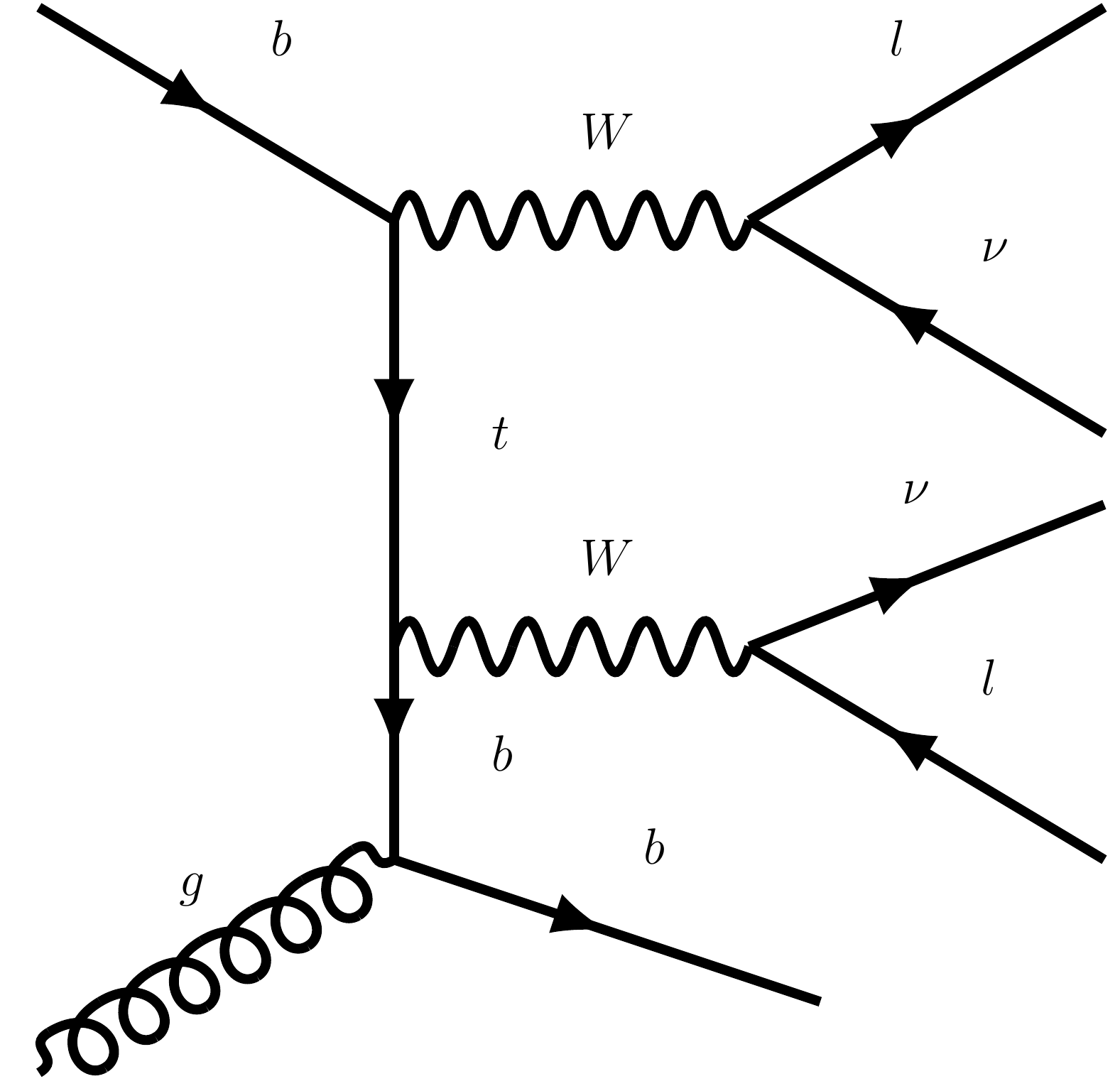}
\caption{Representative leading order SM diagrams for $pp\to b l^+\nu l^- \bar \nu$.} \label{smdiagrams}
\end{figure}

\begin{figure}[h!]
\vspace{0.5cm}
\includegraphics[scale=0.25]{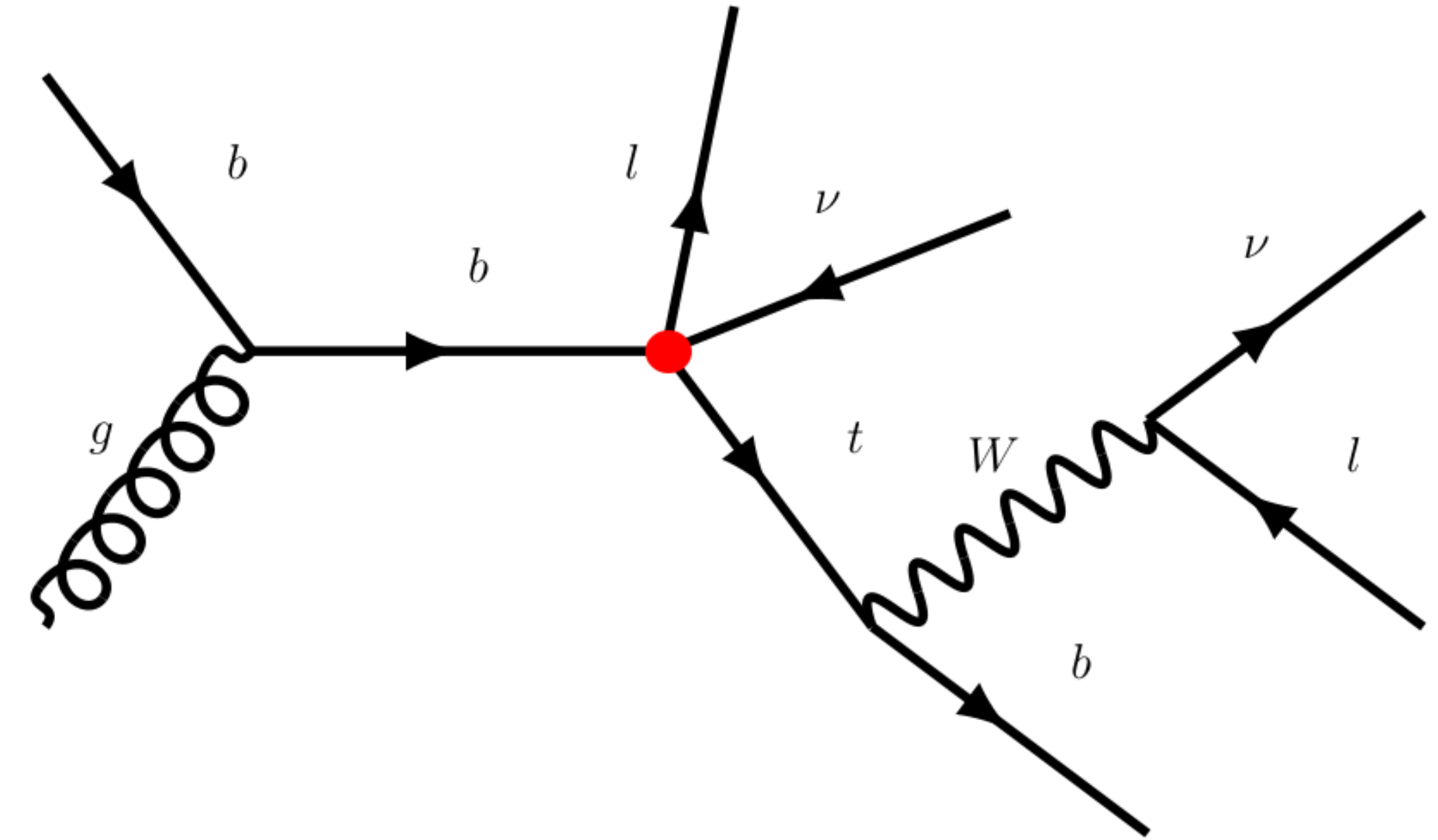} 
\includegraphics[scale=0.25]{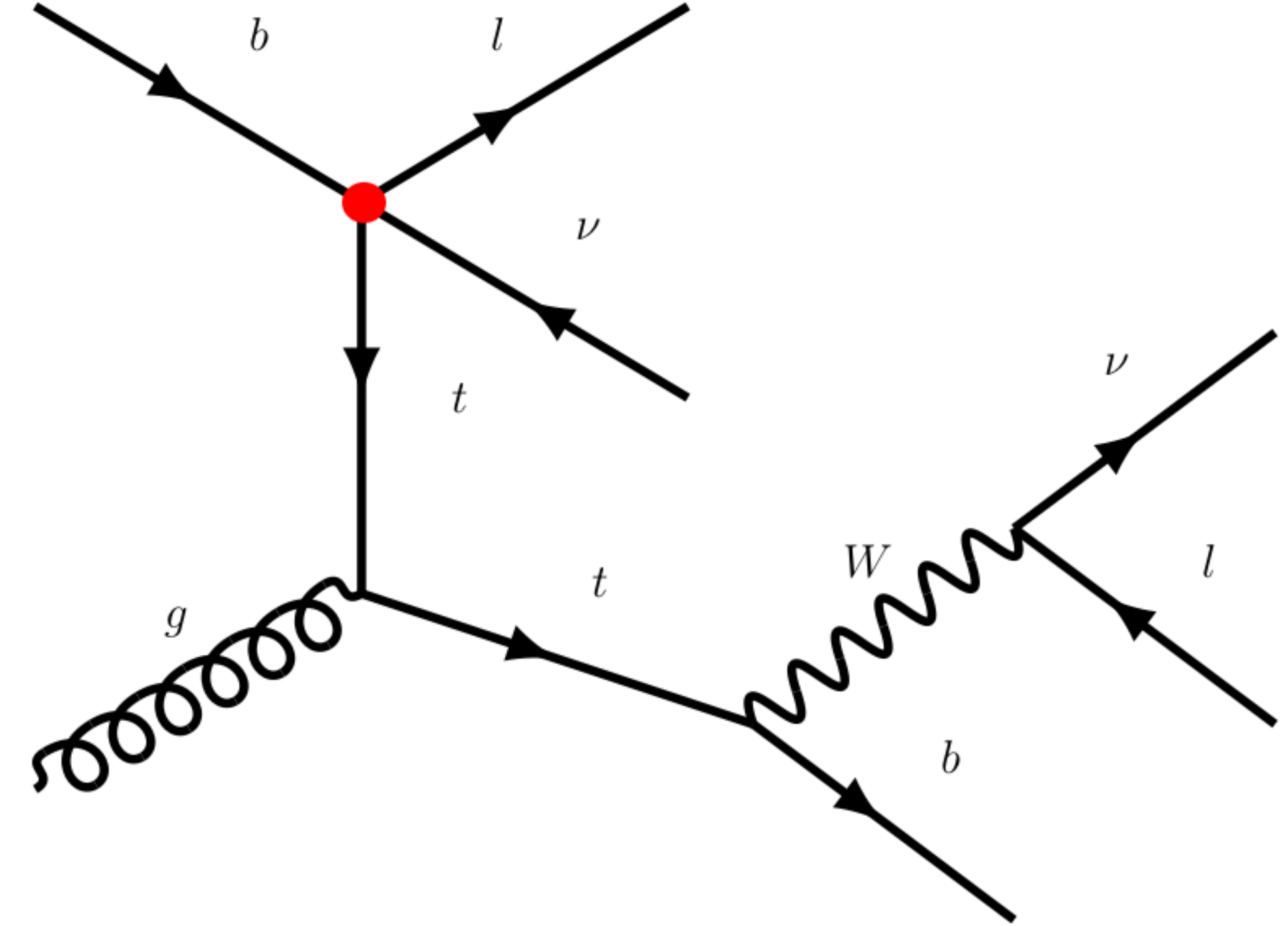} 
\includegraphics[scale=0.25]{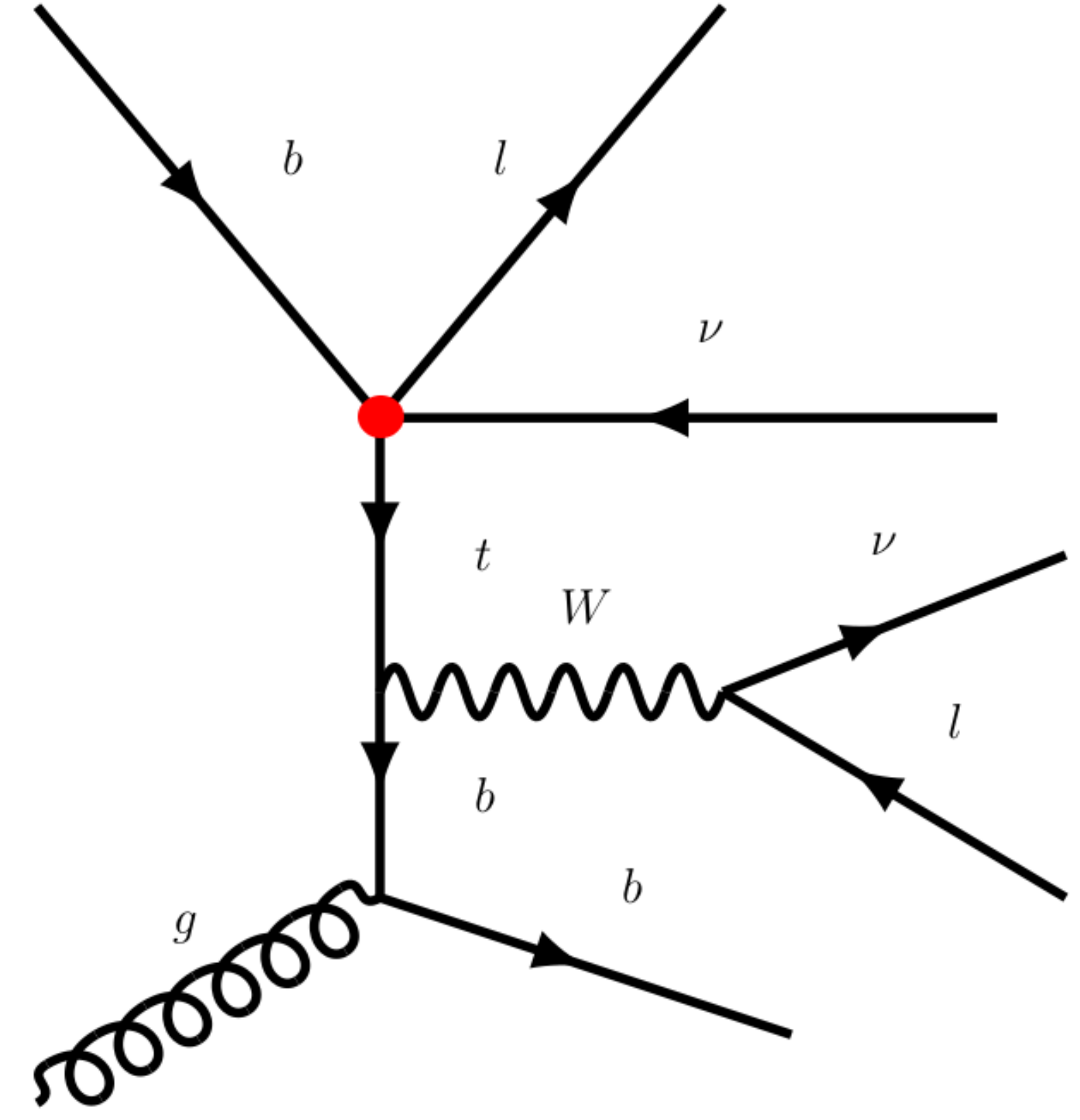}
\caption{Representative leading order EFT diagrams for $pp\to b l^+\nu l^- \bar \nu$. The four fermion interactions are shown with a red dot.} \label{eftdiagrams}
\end{figure}

\begin{figure}[h!]
\vspace{0.5cm}
\includegraphics[scale=0.25]{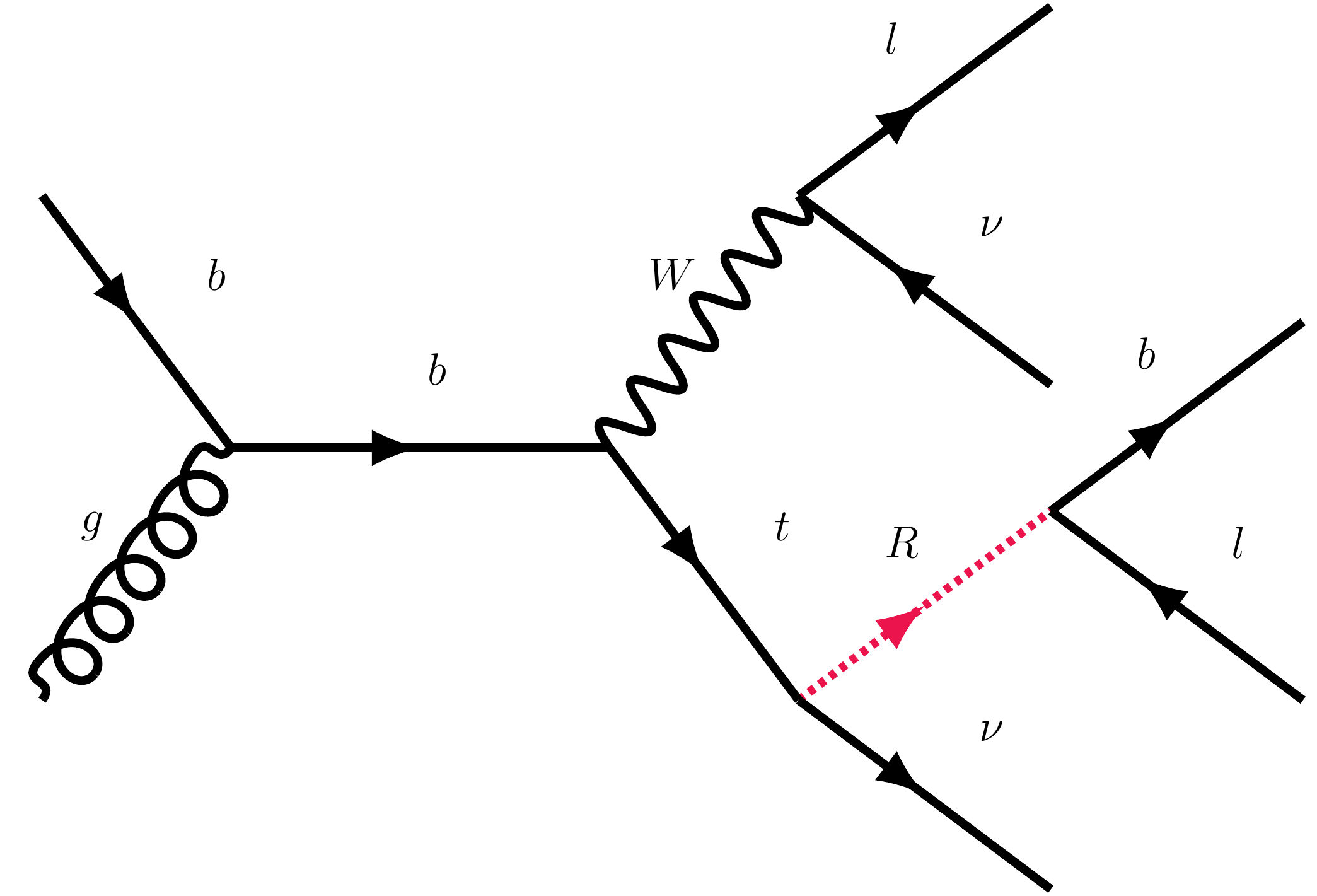} 
\includegraphics[scale=0.25]{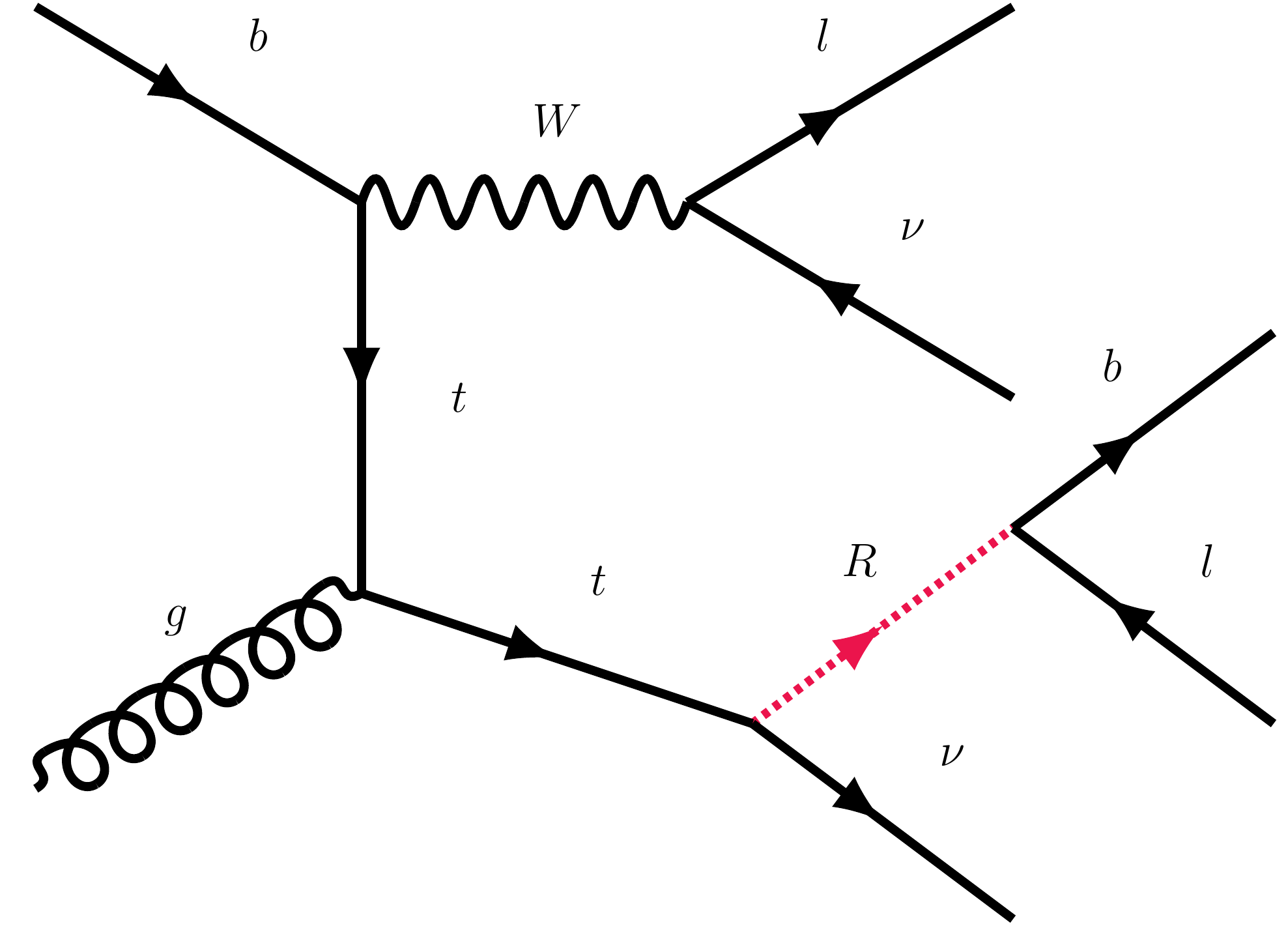} 
\includegraphics[scale=0.25]{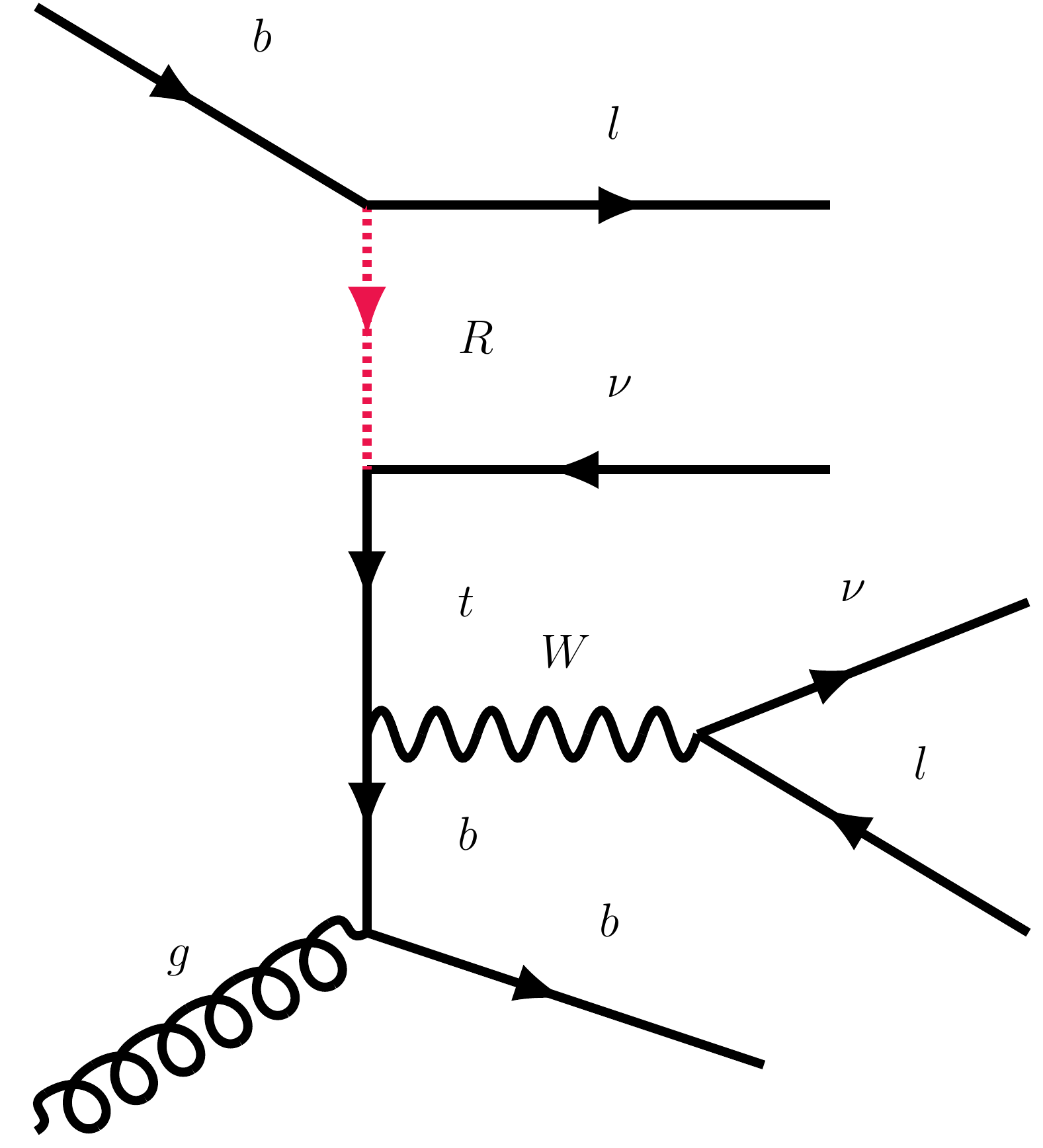}
\caption{Representative leading order LQ model diagrams for $pp\to b l^+\nu l^- \bar \nu$. The LQ lines are shown in red.} \label{lqdiagrams}
\end{figure}

\begin{figure}[h!]
\vspace{0.5cm}
\includegraphics[scale=0.25]{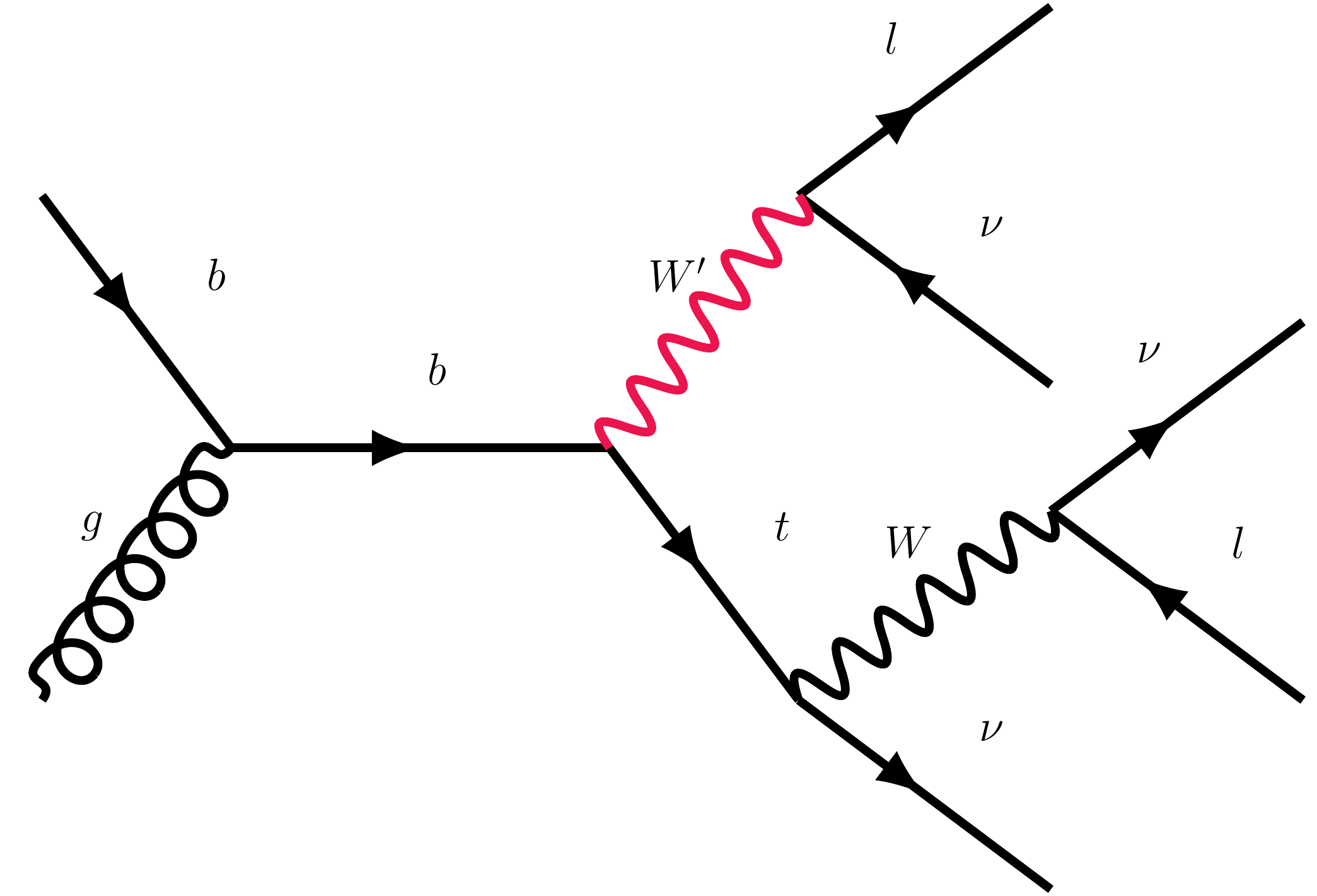} 
\includegraphics[scale=0.25]{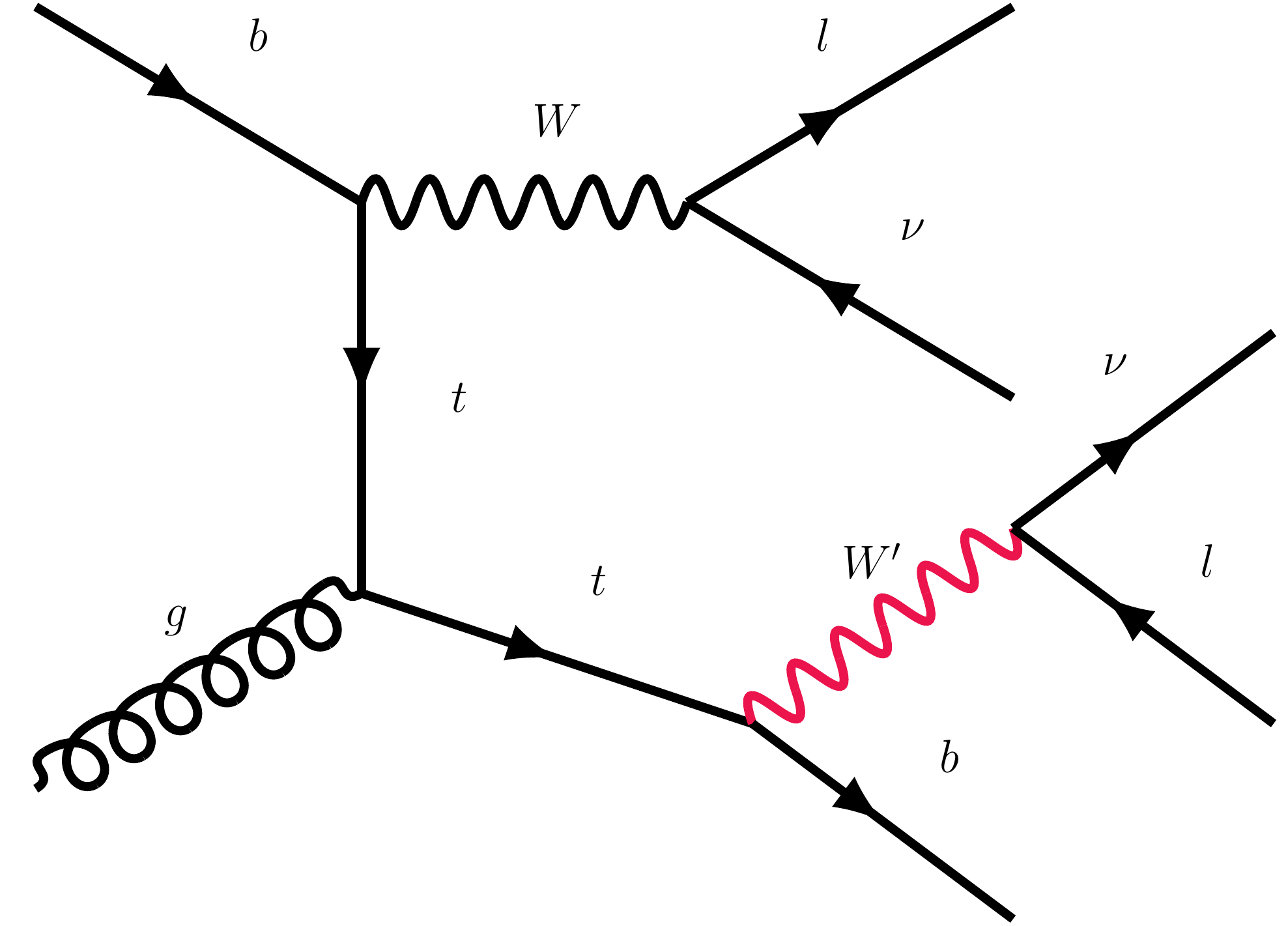} 
\includegraphics[scale=0.25]{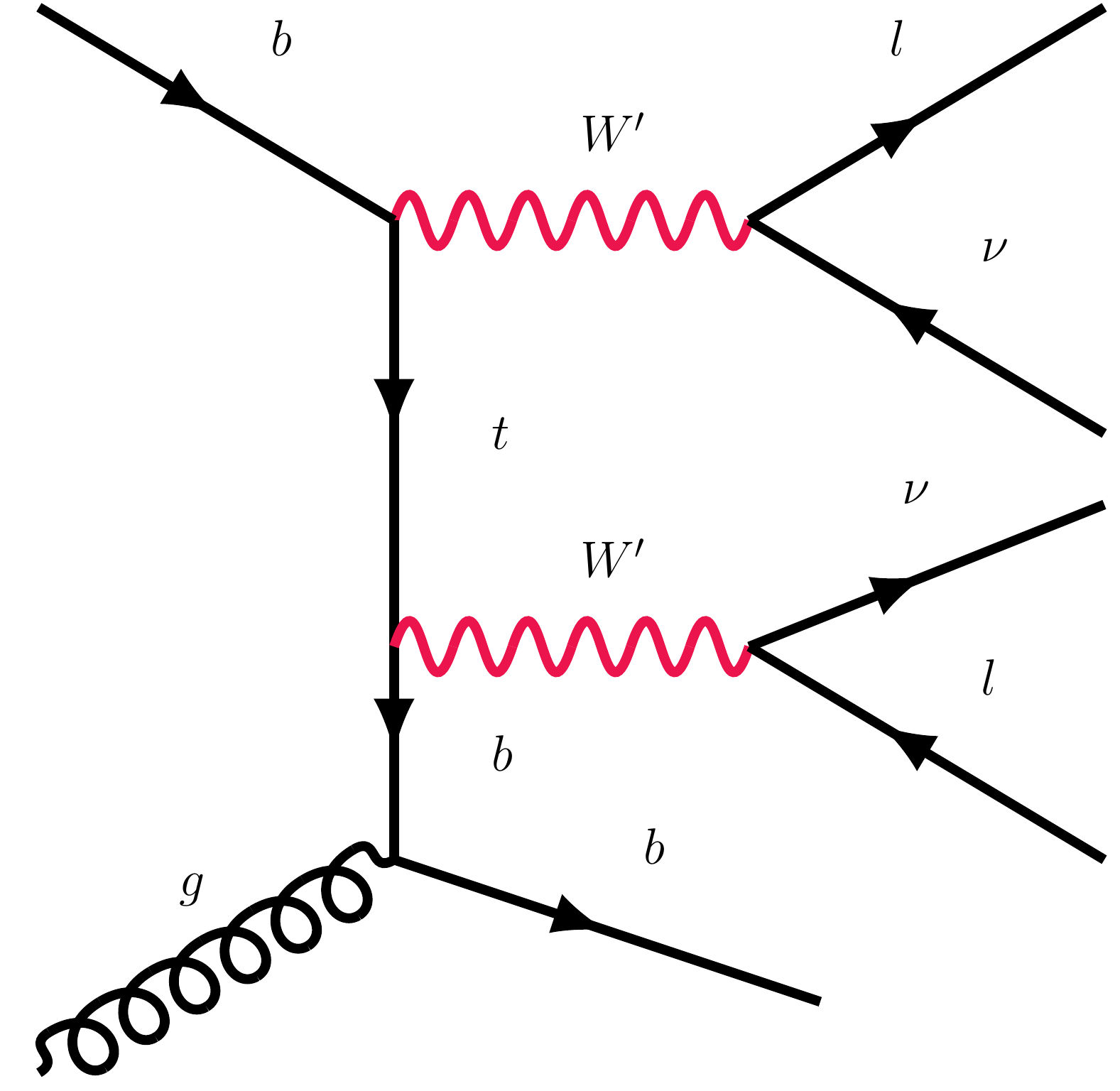}
\caption{Representative leading order $W'$ model diagrams for $pp\to b l^+\nu l^- \bar \nu$. The $W'$ lines are shown in red.} \label{wpdiagrams}
\end{figure}

Here we describe the simulation framework used to compute our results.
In the case of EFT, we implement the operators of Eq.~\eqref{eftops} in \texttt{FeynRules2.0}~\cite{Alloul:2013bka} and generate the corresponding \texttt{UFO} modules to be used for event simulation.
For the scalar leptoquark and $W'$ models of Eq.~\eqref{lagrlq} and \eqref{lagrwp} we use the publicly available UFO modules, respectively \cite{Dorsner:2018ynv} and \cite{wprimeufo}. 
For each BSM model, we simulate $pp\to b l^+\nu l^- \bar \nu$ events at tree-level, where $l=e,\mu$ and $\nu=\nu_e,\nu_\mu$, at center of mass energy $\sqrt{s}=13$ TeV using \texttt{MadGraph5\_aMC@NLO}~\cite{Alwall:2014hca}. In Fig.~\ref{smdiagrams} representative leading order SM diagrams are shown. In Figs.~\ref{eftdiagrams},~\ref{lqdiagrams} and~\ref{wpdiagrams} representative leading order diagrams are shown for EFT, LQ and $W'$ models, respectively. 

These events are subsequently showered with \texttt{PYTHIA8}~\cite{Sjostrand:2007gs,Sjostrand:2006za}. Then, jet clustering is performed using \texttt{FastJet}~\cite{Cacciari:2011ma}, implementing the anti-kt algorithm~\cite{Cacciari:2008gp} with R=0.4. 
Finally, events are selected according to the requirements of the ATLAS analysis~\cite{Aaboud:2017qyi}, namely events must 
\begin{itemize}
\item contain exactly one jet with $p_T>$ 25 GeV and $|\eta|<2.5$ which is also $b$-tagged\footnote{The analysis in~\cite{Aaboud:2017qyi} unfolds the effects of $b$-tagging.}  (the $b$-tagging is performed by requiring that
a parton-level $b$-quark is inside the jet); 
\item contain exactly two oppositely charged leptons ($e$ or $\mu$) with $p_T>$  20 GeV and $|\eta|<2.5$.
\end{itemize}
Because the results of~\cite{Aaboud:2017qyi} are presented unfolded from data, this analysis chain is sufficient to compare to the data. 

With the events that pass the selection criteria, we construct the following normalized differential distribution as defined in the ATLAS analysis~\cite{Aaboud:2017qyi}:
\begin{itemize}
\item the energy of the system of the two leptons and the $b$-jet, $E(llb)$;
\item the mass of the two leptons and the $b$-jet, $m(llb)$;
\item the transverse mass of the leptons, the $b$-jet and the neutrinos, defined to be
\be 
m_T(ll\nu\nu b)=\sqrt{\left(\sum_{i=l_1,l_2,b}p_{T_i} +p_{T_{\rm miss}}\right)^2-\left(\sum_{i=l_1,l_2,b}\overrightarrow{p_{T_i}} +\overrightarrow{p}_{T_{\rm miss}}\right)^2}
\ee
where $p_{T_i}=|\overrightarrow{p_{T_i}}|$ and $\overrightarrow{p}_{T_{\rm miss}}$ is the missing transverse momentum.
\end{itemize}
The ATLAS collaboration actually measures a total of six normalized differential distribution, but we focus on these three because they show better sensitivity to our BSM models. Table~\ref{measdiff} shows a summary of the ATLAS measurements of the normalised differential cross-sections considered in our analysis, with uncertainties shown as percentages.
\begin{table}[h!]
\begin{center}
\vspace{1cm}
\begin{tabular}{|c| c c c c c c |} 
\hline 
$E(llb)$ bin [GeV] & [50,175] & [175,275]& [275,375] & [375,500] & [500,700] & [700,1200]\tabularnewline
$(1/\sigma) d\sigma/dE $ [GeV$^{-1}$] & 0.000597 & 0.00322 & 0.00185 & 0.00135 & 0.000832 & 0.000167 \tabularnewline
Total uncertainty [\%] & 38 & 18 & 22 & 56 & 53 & 45 \tabularnewline
\hline 
$m(llb)$ bin [GeV] & [0,125] & [125,175]& [175,225] & [225,300] & [300,400] & [400,1000]\tabularnewline
$(1/\sigma) d\sigma/dm $ [GeV$^{-1}$] & 0.00051 & 0.00533 & 0.00538 & 0.00242 & 0.000949 & 0.000208 \tabularnewline
Total uncertainty [\%] & 43 & 20 & 21 & 26 & 30 & 34 \tabularnewline
\hline 
$m_T(ll\nu\nu b)$ bin [GeV] & [50,275] & [275,375]& [375,500] & [500,1000] & & \tabularnewline
$(1/\sigma) d\sigma/dm_T $ [GeV$^{-1}$] & 0.0033 & 0.00123 & 0.000856 & 5.51$\times 10^{-5}$ & & \tabularnewline
Total uncertainty [\%] & 11 & 48 & 43 & 55 & & \tabularnewline
\hline 
\end{tabular}
\caption{\label{measdiff} Summary of the measured normalised differential cross-sections as in~\cite{Aaboud:2017qyi}, with uncertainties shown as percentages.} 
\end{center}
\end{table}

The BSM computation is performed at tree-level, and in order to compare our predictions with the ATLAS results, we need approximate next-to-leading order (NLO) precision. Therefore, we rescale each distribution by appropriate bin-dependent $k$-factors that we estimate by taking the ratio of our SM predictions computed at NLO in $\alpha_s$ to the tree-level SM predictions. The SM NLO computation is performed in \texttt{MadGraph5\_aMC@NLO}~\cite{Alwall:2014hca}
by applying the diagram removal procedure~\cite{Frixione:2008yi} where one removes all diagrams in the NLO real emission amplitudes that are doubly $t$-resonant. 

The resulting $k$-factors for $m(llb)$ and $m_T(ll\nu\nu b)$ distributions are shown in Table~\ref{kfactors}.\footnote{We only use the $E(llb)$ observable when computing ratios in Section~\ref{sec:ratios} where NLO effects cancel, so we have not computed the $k$-factor for this observable. } We can see that these $k$-factors vary between 0.7 to 1.1 depending on the measured quantity and the bin. We will apply these $k$-factors to our tree-level predictions of EFT, scalar leptoquark, and $W'$ models.

\begin{table}[h!]
\begin{center}
\vspace{1cm}
\begin{tabular}{|c| c c c c c c |} 
\hline 
$m(llb)$ bin [GeV] & [0,125] & [125,175]& [175,225] & [225,300] & [300,400] & [400,1000]\tabularnewline
$(1/\sigma) d\sigma/dm $ [GeV$^{-1}$] NLO & 0.00063 & 0.00513 & 0.00548 & 0.00296 & 0.00107 & 0.000104 \tabularnewline
$(1/\sigma) d\sigma/dm $ [GeV$^{-1}$] LO & 0.00061 & 0.00494 & 0.00536 & 0.00302 & 0.00112 & 0.000117 
\tabularnewline
$k$-factor & 1.03 & 1.04 & 1.02 & 0.98 & 0.96 & 0.89 \tabularnewline
\hline 
$m_T(ll\nu\nu b)$ bin [GeV] & [50,275] & [275,375]& [375,500] & [500,1000] & & \tabularnewline
$(1/\sigma) d\sigma/dm_T $ [GeV$^{-1}$] NLO & 0.00315 & 0.00205 & 0.000482 & 5.17$\times 10^{-5}$ & & \tabularnewline
$(1/\sigma) d\sigma/dm_T $ [GeV$^{-1}$] LO & 0.00290 & 0.00233 & 0.000628 & 7.31$\times 10^{-5}$ & & 
\tabularnewline
$k$-factor & 1.09 & 0.88 & 0.77 & 0.71 & & \tabularnewline
\hline 
\end{tabular}
\caption{$k$-factors for $m(llb)$ and $m_T(ll\nu\nu b)$ that we estimate by comparing the distributions of NLO with tree-level SM events generated with \texttt{MadGraph5\_aMC@NLO} and showered with \texttt{PYTHIA8}.\label{kfactors}} 
\end{center}
\end{table}

\section{Analysis and Results}
\label{sec:results}
In this section we give our results, focusing on computing 95\% CL limits using current data and projecting limits using future data.
In order to determine the limits on the parameters of our BSM models (EFT coefficients and simplified model couplings), we implement a simple chi-squared analysis. For each differential measurement of the observable  $x$ the following reduced $\chi^{2}$ function
\begin{equation}
\chi^{2}=\frac{1}{N_x-1}\sum_{i=1}^{N_x}\frac{\left[\frac{1}{\sigma}\left(\frac{d\sigma}{dx}\right)_i-\frac{1}{\sigma}\left(\frac{d\sigma}{dx}\right)_i^{\rm exp}\right]^{2}}{(\delta \sigma_i^{{\rm exp}})^2}\, ,
\label{chisq}
\end{equation}
where $N_x$ is the number of bins used in the measurement of the differential distribution of the observable $x$ (see Table~\ref{measdiff}), $\frac{1}{\sigma}\left(\frac{d\sigma}{dx}\right)_i$ is the normalized differential distribution in the $i$-th bin computed in the presence of new physics, $\frac{1}{\sigma}\left(\frac{d\sigma}{dx}\right)_i^{\rm exp}$ is the measured normalized differential distribution and $\delta \sigma_i^{{\rm exp}}$ is the total experimental uncertainty (stat+syst) in the $i$-th bin which is reported as percentage of the measured value in Table~\ref{measdiff}. We do not include any theoretical uncertainties which are negligible with respect to the experimental ones. Moreover, no correlation matrix among different bin uncertainties is considered since this information has not been made publicly available yet by the ATLAS experimental collaboration. For each BSM model, we scan over the model parameters and compute the $\chi^{2}$ function of Eq.~\eqref{chisq}. Parameters are considered excluded at 95\% CL if $\chi^{2}>2.2$ for $x=E(llb)$ or $x=m(llb)$. On the other hand, parameters are considered excluded at 95\% CL if $\chi^{2}>2.6$ for $x=m_T(ll\nu \nu b)$ because of the smaller number of bins.

\subsection{EFT: current bounds at 36.1 fb$^{-1}$}
\label{sec:eft_current}
\begin{figure}[h!]
\vspace{0.5cm}
\includegraphics[scale=0.50]{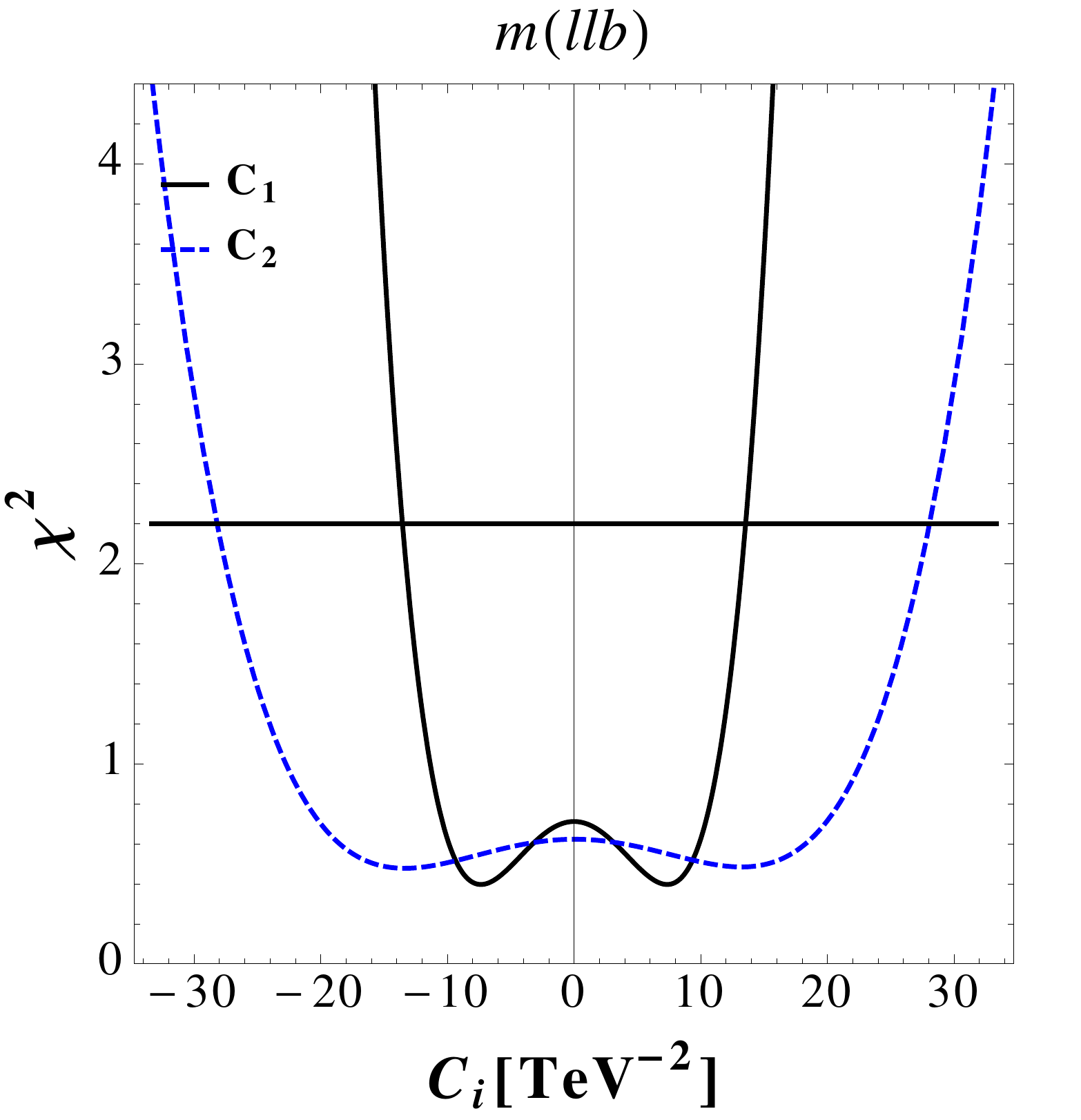} 
\includegraphics[scale=0.50]{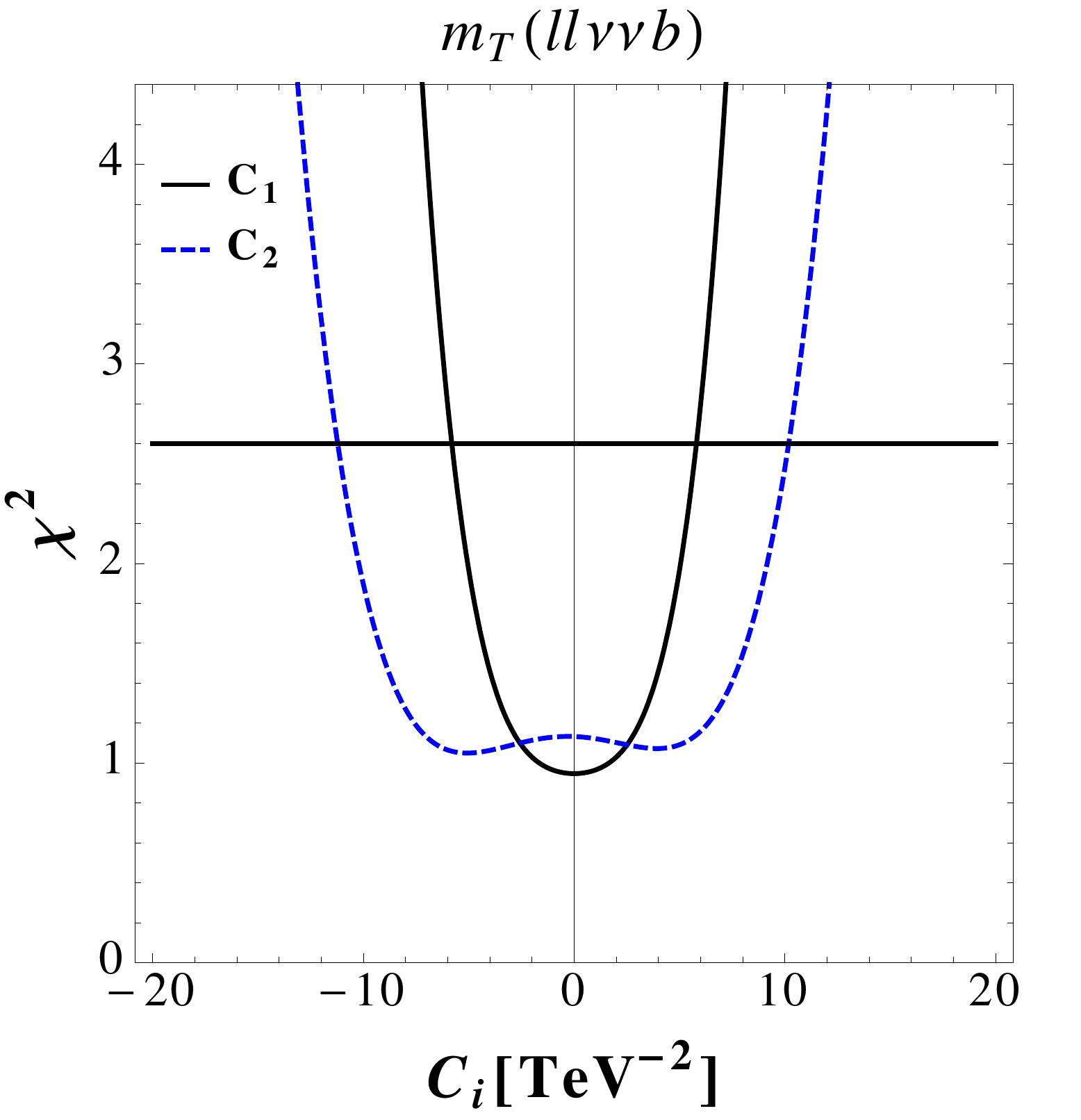} 
\caption{$\chi^2$ as function of $C_1$ (solid black curve) and $C_2$ (dashed blue curve) obtained by using the $m(llb)$ (left plot) and $m_T(ll\nu \nu b)$ (right plot) differential distribution measurements at 36.1 fb$^{-1}$.} \label{boundeftcurrent}
\end{figure}

In the EFT case, where BSM phyiscs is parametrized by the effective operators in Eq.~\eqref{eftops}, we consider two possible configurations for the operator coefficients. Configuration (i) corresponds to the {\it LQ-like} case in which the coefficient of the first operator in Eq.~\eqref{eftops} is set to zero and the other two are taken to be proportional, as described in Eq.~\eqref{case1}. In this case, BSM effects  are parametrized by a single effective coefficient that we called $C_1$. Configuration (ii) corresponds to the {\it $W'$-like} case in which just the first operator coefficient is non-zero and is described by Eq.~\eqref{case2}. In this case, the BSM effects are parametrized by a single effective coefficient that we called $C_2$. In Fig.~\ref{boundeftcurrent} we present current bounds at 95\% CL on the EFT coefficients $C_1$ and $C_2$ obtained by using the $m(llb)$ (left plot) and $m_T(ll\nu \nu b)$ (right plot) differential distribution measurements at 36.1 fb$^{-1}$~\cite{Aaboud:2017qyi}. More details on the simulation and fitting procedure can be found in Appendix~\ref{app:fitting}. The solid black curve represents the $\chi^2$ as function of $C_1$, while the dashed blue curve represents the $\chi^2$ as function of $C_2$.

From Fig.~\ref{boundeftcurrent} we can see that the $m_T(ll\nu \nu b)$ differential distribution measurement turns out to be more sensitive to the EFT operators and it provides the following constraints
\be 
|C_1|\lesssim 5\times  \,{\rm TeV}^{-2}\qquad\qquad |C_2|\lesssim 10 \,{\rm TeV}^{-2} \; .
\label{eq:c_current}
\ee
Using the relations in Eq.~\eqref{case1} and \eqref{case2} and assuming an ${\cal O}(1)$ value for the $c_i$ coefficients we can translate those bounds into a bound on the BSM scale
\be
\Lambda \gtrsim 300{\rm -}400\, {\rm GeV} 
\label{eq:lam_current}
\ee
If instead we use the matching condition of Eq.~\eqref{match1} and \eqref{match2} we get
\be \label{eftlinelq}
M_R\gtrsim g\times 150 \, {\rm GeV}
\ee
and 
\be \label{eftlinewp}
M_{W'}\gtrsim  k_L\times 100\, {\rm GeV}
\ee
respectively. 

We see from comparing the energy scales for these bounds to the high energy bins of our observables, that our limits are outside the regime of validity of the EFT for perturbative couplings. This is particularly a problem for differential analyses like this one, where the strongest discrimination power comes from the high energy bins. As we will see in the next subsection, with additional data the limits become stronger, but they are still not generically within the regime of validity of an EFT analysis. Therefore, to get a strictly valid analysis, it must be done in the framework of a renormalizable model,\footnote{The massive $W'$ still needs a mechanism to give mass the vector, but that does not necessarily affect our analysis. } which we do in Sections~\ref{sec:lq_current}--\ref{sec:wp_future}.

\subsection{EFT: expected bounds at 300 and 3000 fb$^{-1}$}
\label{sec:eft_future}
In this section we present the expected bounds on $C_1$ and $C_2$ at 300 and 3000 fb$^{-1}$ obtained by using the same differential distributions. We consider two possible scenarios for how uncertainties will scale with additional data. The first is an optimistic one in which the systematic uncertainties are assumed to scale in accordance to the statistical uncertainties as ${\cal L}^{-1/2}$. The second scenario is a pessimistic one in which the systematic uncertainties remain unchanged with respect to the current uncertainties of~\cite{Aaboud:2017qyi}. In the reduced $\chi^2$ computation we fix the measured value to coincide with the SM theoretical prediction. More details on the simulation and fitting procedure can be found in Appendix~\ref{app:fitting}.
In Fig.~\ref{boundeftexpected} the reduced $\chi^2$ as function of the effective operator coefficients $C_1$ (left plot) and $C_2$ (right plot) is shown for the different scenarios considered at 300 and 3000 fb$^{-1}$: solid lines represent the pessimistic scenario for future systematic errors, while dashed lines represent the optimistic one. In both plots the $m_T$ distribution has been used to derive the expected bounds since it provides the best sensitivity.
\begin{figure}[h!]
\vspace{0.5cm}
\includegraphics[scale=0.43]{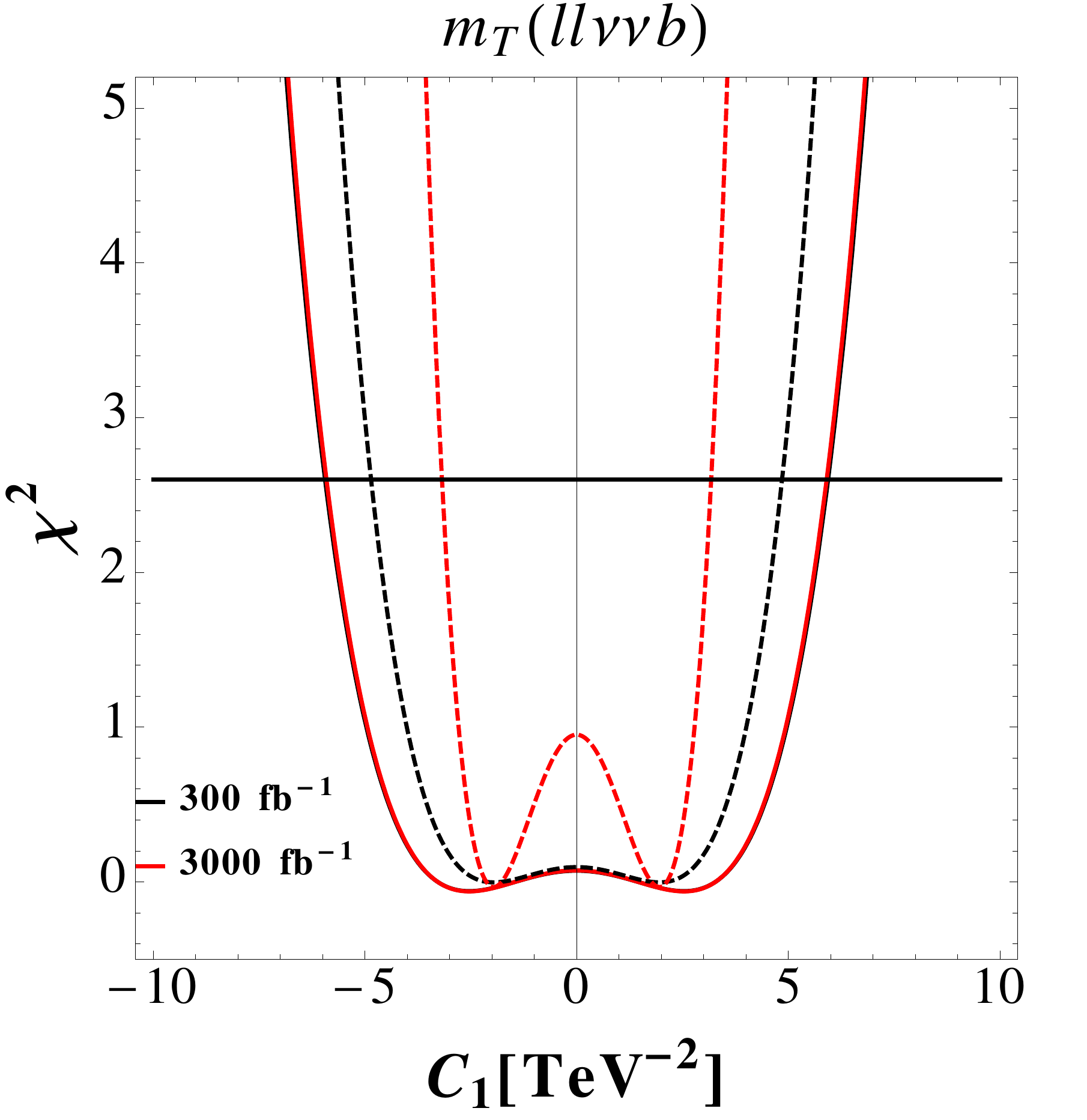} 
\includegraphics[scale=0.425]{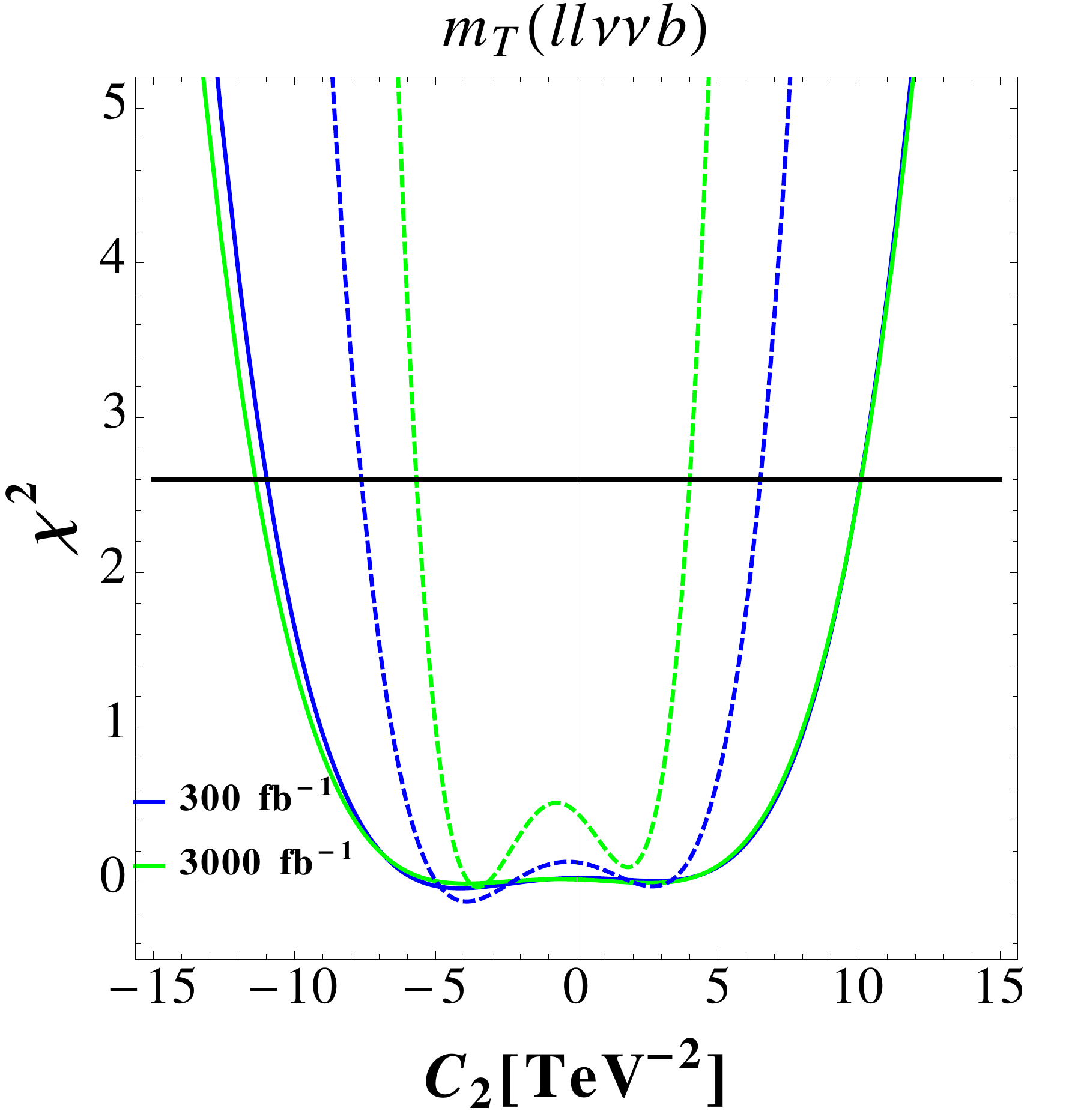} 
\caption{
Expected behavior of the reduced $\chi^2$ as function of the effective operator coefficients $C_1$ (left plot) and $C_2$ (right plot) obtained using the $m_T$ distribution. In the left plot, the black solid (dashed) curve corresponds to the pessimistic (optimistic) scenario for 300 fb$^{-1}$, while the red solid (dashed) curve corresponds to the pessimistic (optimistic) scenario for 3000 fb$^{-1}$. In the right plot, the blue solid (dashed) curve corresponds to the pessimistic (optimistic) scenario for 300 fb$^{-1}$, while the green solid (dashed) curve corresponds to the pessimistic (optimistic) scenario for 3000 fb$^{-1}$.} \label{boundeftexpected}
\end{figure}
Looking at the plots we can see that the pessimistic scenarios do not provide any improvement on the limits regardless to the luminosity, this is due to the fact that the uncertainty is dominated by systematic errors.
Looking at the most optimistic scenario at 3000 fb$^{-1}$ we have instead the following improvement on the bounds
\be 
|C_1|\lesssim 3 \,{\rm TeV}^{-2}\qquad\qquad |C_2|\lesssim 5\,{\rm TeV}^{-2}  \; .
\ee
Using the relations in Eq.~\eqref{case1} and \eqref{case2} and assuming an ${\cal O}(1)$ value for the $c_i$ coefficients we can translate those bounds into a bound on the BSM scale
\be
\Lambda \gtrsim 400{\rm -}600\, {\rm GeV} 
\ee
which is higher than Eqs.~\eqref{eq:c_current} and~\eqref{eq:lam_current}, but still well below the invariant mass the high energy events in this analysis. 

\subsection{Scalar leptoquark model: current bounds at 36.1 fb$^{-1}$}
\label{sec:lq_current}
\begin{figure}[h!]
\vspace{0.5cm}
\includegraphics[scale=0.7]{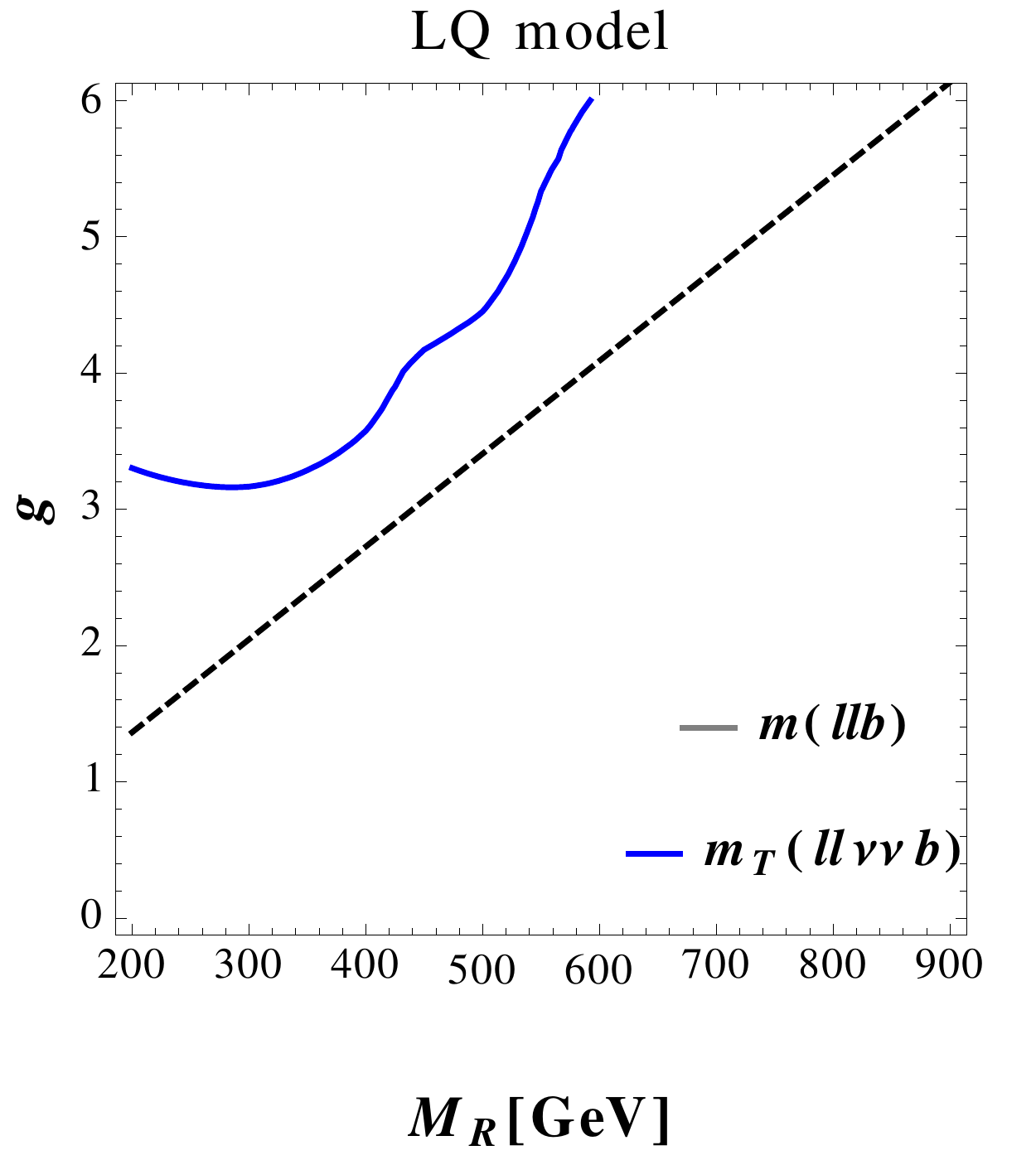} 
\caption{95\% CL bounds in the $M_R$ vs $g$  plane obtained by using the $m_T(ll\nu \nu b)$ (blue curve) differential distribution measurement at 36.1 fb$^{-1}$. The dashed black line represnts the EFT bound of Eq.~\eqref{eftlinelq}. \label{boundlq}}
\end{figure}

Here we consider the scalar leptoquark model of Section IIIA. 
This model is characterized by two parameters, the coupling $g$ and the mass of the leptoquark $M_R$.\footnote{The relative sign of the Yukawa couplings $z_{13}$ and $y_{31}$ is in principle observable, but it does not affect our analysis.}  In Fig.~\ref{boundlq} we present 95\% CL bounds in the $M_R$ vs $g$ plane obtained by using the $m(llb)$ (grey curve) and $m_T(ll\nu \nu b)$ (blue curve) differential distribution measurements and considering same sign Yukawa couplings $z_{13}$ and $y_{31}$. More details on the simulation and fitting procedure can be found in Appendix~\ref{app:fitting}.
We consider values for the leptoquark mass bigger than 200 GeV in order to avoid top quark decay to an on-shell leptoquark, $t\to R l$, which would strongly constrain that region of the parameter space. We also note that recasts such as those in~\cite{Diaz:2017lit} exclude a significant portion of the considered parameter space, and a dedicated analysis with current LHC data could like exclude even more.
We allow our coupling $g$ to be relatively large, near and possibly beyond the boundary of where perturbation theory is valid. At high end of the range we consider, higher order effects become very important. Our simulations will remain at leading order except when indicated otherwise, but we note that theoretical uncertainties on our limits at large couplings are significant.

The excluded regions in Fig.~\ref{boundlq}  lie above the curves. Here the best sensitivity is still obtained by considering the $m_T$ differential distribution. The plot obtained by taking opposite sign Yukawa couplings $z_{13}$ and $y_{31}$ gives identical exclusion regions because this model does not interfere with the SM at tree level.
In Fig.~\ref{boundlq}, for comparison we also show the EFT bound of Eq.~\eqref{eftlinelq}, which corresponds to a straight line in this plane. In accordance with our conclusion in Sec.~\ref{sec:eft_current}, we see that the EFT bound is everywhere stronger than the bound in the LQ model, confirming that the EFT analysis is not strictly valid. The reason the EFT bounds are stronger is because the dominant diagrams in the full theory are $t$-channel, such as the right diagram in Fig.~\ref{lqdiagrams}. Because $t<0$, the effect of the EFT will be larger than that from the full theory, so the extracted bound will be stronger.

\subsection{Scalar leptoquark model: expected bounds at 300 and 3000 fb$^{-1}$}
\label{sec:lq_future}
In this section we present the expected bounds at 300 and 3000 fb$^{-1}$ in the $M_R$ vs $g$ plane obtained by using the same differential distributions of the previous section and considering same sign Yukawa couplings $z_{13}$ and $y_{31}$. As in Sec.~\ref{sec:eft_future}, we consider two possible scenarios: an optimistic one in which the systematic uncertainties are assumed to scale in accordance with the statistical uncertainties and a pessimistic one in which the systematic uncertainties remain unchanged with respect to the current uncertainties of~\cite{Aaboud:2017qyi}. In the reduced $\chi^2$ computation we fix the measured value to coincide with the SM theoretical prediction.  More details on the simulation and fitting procedure can be found in Appendix~\ref{app:fitting}.
In Fig.~\ref{bexplq} the 95\% CL expected exclusion regions are shown for the different scenarios considered: the left plot makes use of the $m(llb)$ distribution while the right plot makes use of the $m_T(ll\nu \nu b)$ distribution.
\begin{figure}[h!]
\vspace{0.5cm}
\includegraphics[scale=0.60]{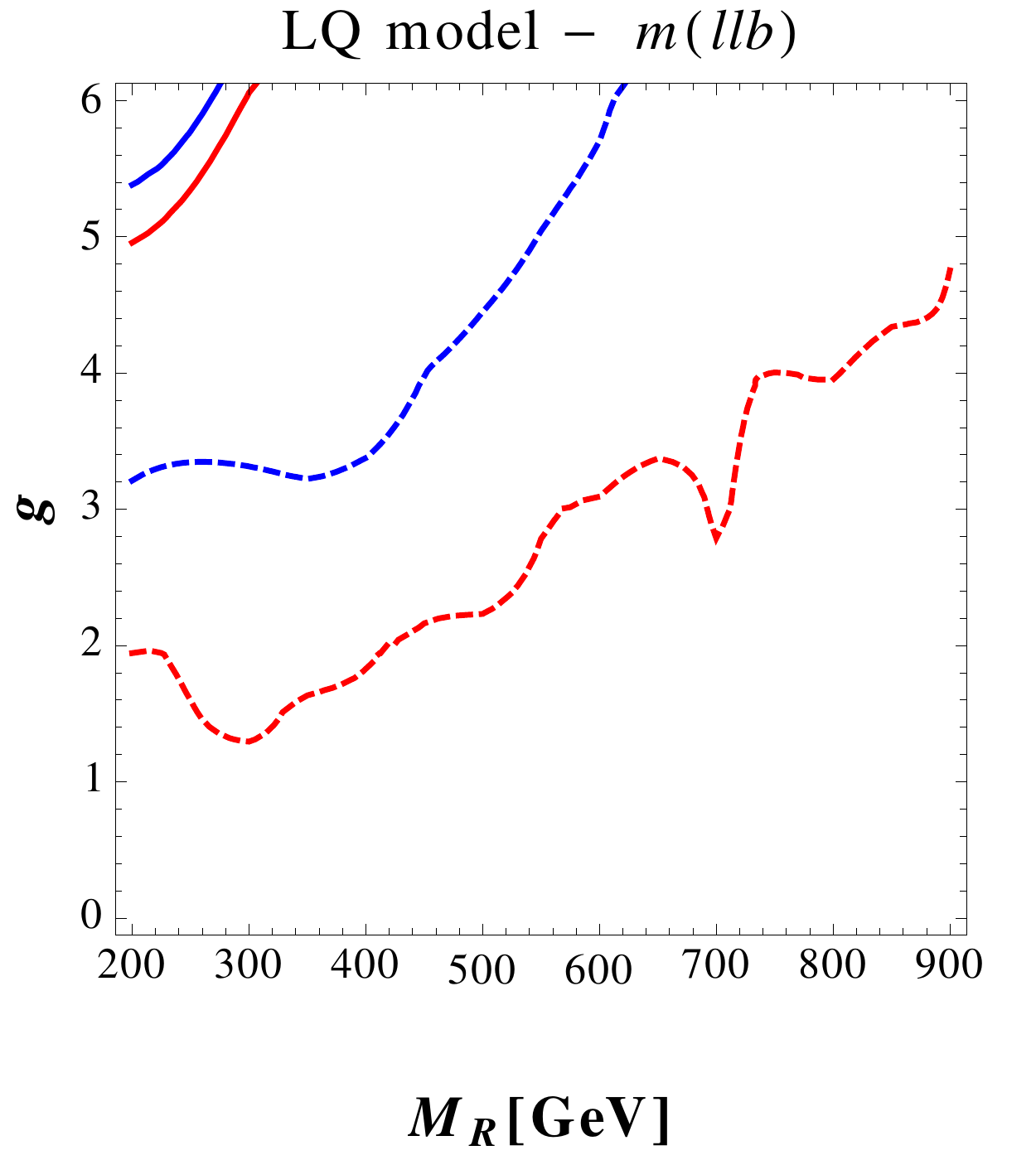} 
\includegraphics[scale=0.60]{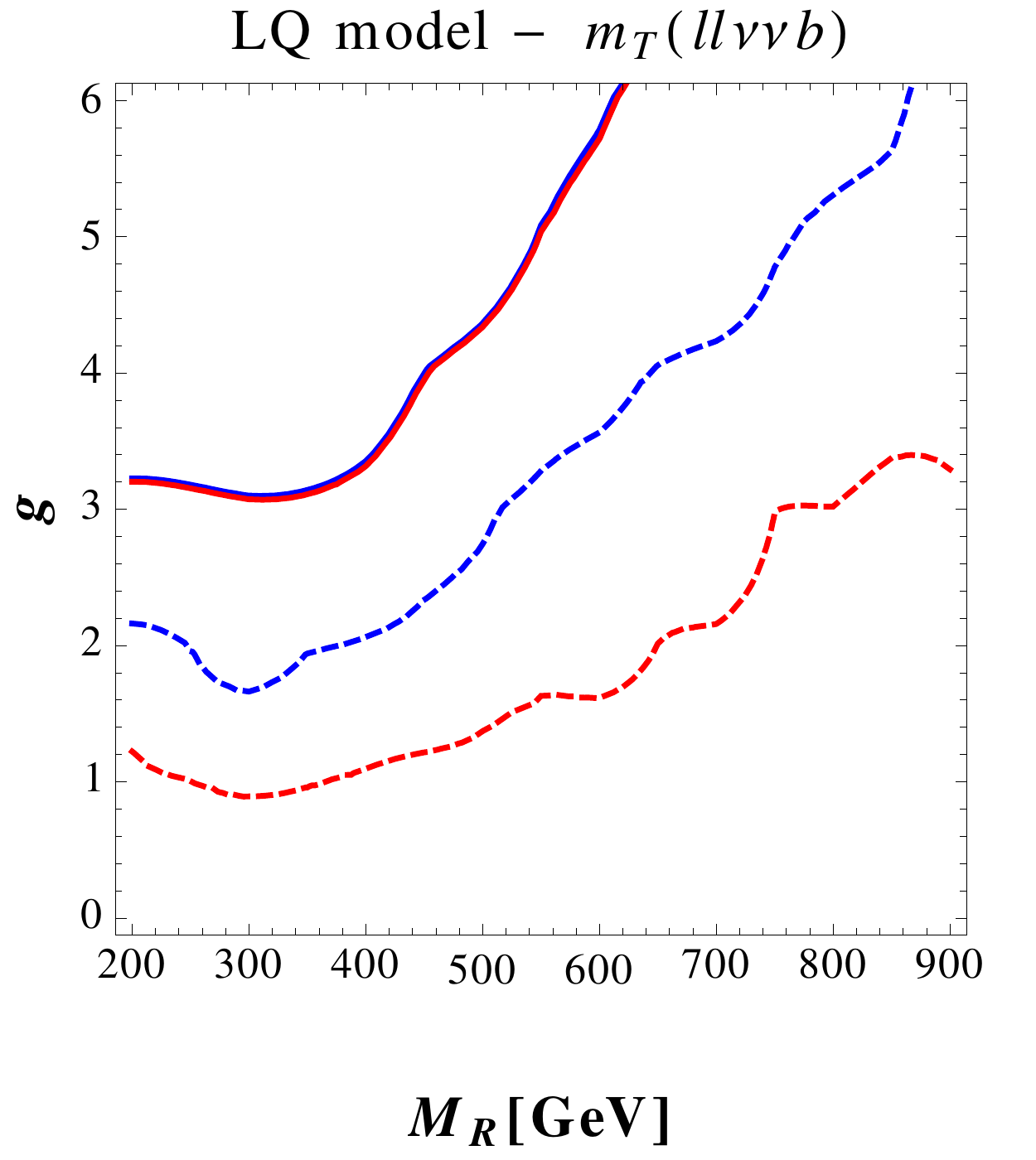} 
\caption{95\% CL expected exclusion regions in the $M_R$ vs $g$ plane: the left plot makes use of the $m(llb)$ distribution while the right plot makes use of the $m_T(ll\nu \nu b)$ distribution. The solid (dashed) blue curves correspond to the pessimistic (optimistic) scenario at 300 fb$^{-1}$ while the solid (dashed) red curves correspond to the pessimistic (optimistic) scenario at 3000 fb$^{-1}$.} \label{bexplq}
\end{figure}
In Fig.~\ref{bexplq} the solid (dashed) blue curves correspond to the pessimistic (optimistic) scenario at 300 fb$^{-1}$ while the solid (dashed) red curves correspond to the pessimistic (optimistic) scenario at 3000 fb$^{-1}$. Looking at the plots we see that $m_T(ll\nu \nu b)$ can provide significant improvements over current data in the optimistic scenario. On the other hand, the pessimistic scenarios do not provide any improvements on the limits. As before, the plots obtained by taking opposite sign Yukawa couplings $z_{13}$ and $y_{31}$ give identical exclusion regions and are not shown. This is due to the fact that the SM-BSM interference contribution is null.

\subsection{$W'$ model: current bounds at 36.1 fb$^{-1}$}
\begin{figure}[h!]
\vspace{0.5cm}
\includegraphics[scale=0.70]{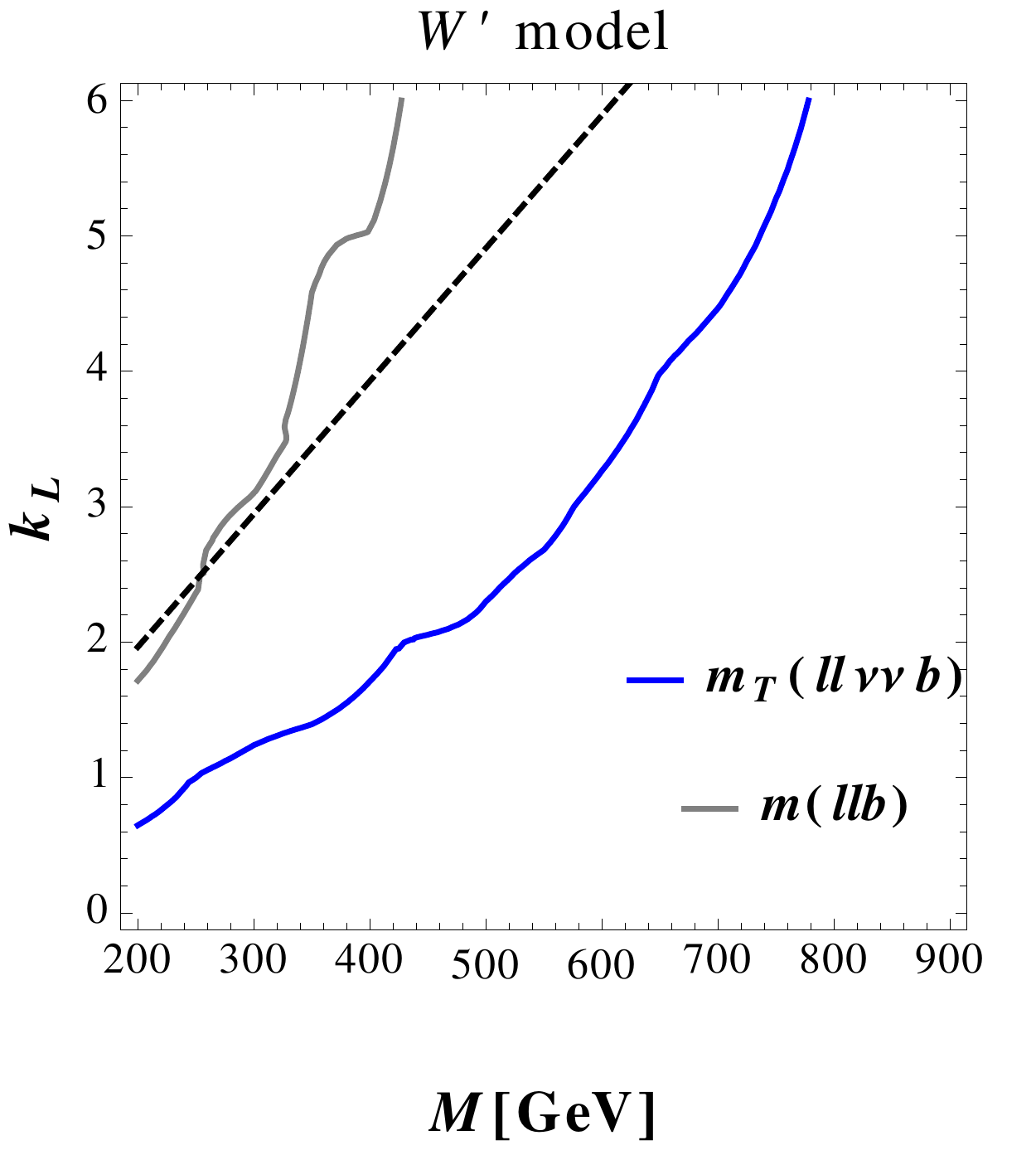} 
\caption{95\% CL bounds in the $M$ vs $k_L$  plane obtained by using the $m(llb)$ (grey curve) and $m_T(ll\nu \nu b)$ (blue curve) differential distribution measurements at 36.1 fb$^{-1}$. The dashed black line represnts the EFT bound of Eq.~\eqref{eftlinewp}.} \label{boundwp}
\end{figure}

Here we consider the $W'$ model of Section IIIB. This model is characterized by two parameters, the coupling rescaling factor $k_L$ and the mass of the $W'$ boson $M$.\footnote{As with the LQ model, the relative sign of the rescaling coefficients  $k_L^l$ and $k_L^q$ is observable, but has very small effects on this analysis.}  In Fig.~\ref{boundwp} we present 95\% CL bounds in the $M$ vs $k_L$  plane obtained by using the $m(llb)$ (grey curve) and $m_T(ll\nu \nu b)$ (blue curve) differential distribution measurements and considering same sign rescaling coefficients  $k_L^l$ and $k_L^q$. More details on the simulation and fitting procedure can be found in Appendix \ref{app:fitting}. 
The excluded regions lie above the curves. We consider values for the $W'$ mass bigger than 200 GeV in order to avoid the top quark decay $t\to W' b$ which would strongly constrain this region of the parameter space. The plots obtained by taking opposite sign rescaling coefficients  $k_L^l$ and $k_L^q$ give similar exclusion limits and are not shown. This is due to the fact that the SM-BSM interference turns out to be very small. 

In Fig.~\ref{boundwp} we show for comparison also the EFT bound of Eq.~\eqref{eftlinewp}, and we see that just as in the LQ case, the bounds differ significantly. In this case, the EFT bounds are \textit{weaker} than from the full theory, as opposed to the LQ case. This is because the dominant diagrams are now $s$-channel like the diagrams in Fig.~\ref{wpdiagrams}, and $0 < s \lesssim M_R$ so the effects in the full theory are larger than the EFT.

\subsection{$W'$ model: expected bounds at 300 and 3000 fb$^{-1}$}
\label{sec:wp_future}
In this section we present the expected bounds at 300 and 3000 fb$^{-1}$ in the $M$ vs $k_L$ plane obtained by using the same differential distributions of the previous section and taking same sign rescaling coefficients  $k_L^l$ and $k_L^q$. We consider the same optimistic and pessimistic scenarios as in Secs.~\ref{sec:eft_future} and~\ref{sec:lq_future}.
In the reduced $\chi^2$ computation, we fix the measured value to coincide with the SM theoretical prediction. More details on the simulation and fitting procedure can be found in Appendix \ref{app:fitting}. 
In Fig.~\ref{bexpwp} the 95\% CL expected exclusion regions are shown for the different scenarios considered: the exclusion regions in the left plot are obtained by using the $m(llb)$ distribution while the exclusion regions in the right plot are obtained from the $m_T(ll\nu \nu b)$ distribution. The plots obtained by taking opposite sign rescaling coefficients  $k_L^l$ and $k_L^q$ give similar exclusion limits and are not shows. This is due to the fact that the SM-BSM interference turns out to be very small.
\begin{figure}[h!]
\vspace{0.5cm}
\includegraphics[scale=0.62]{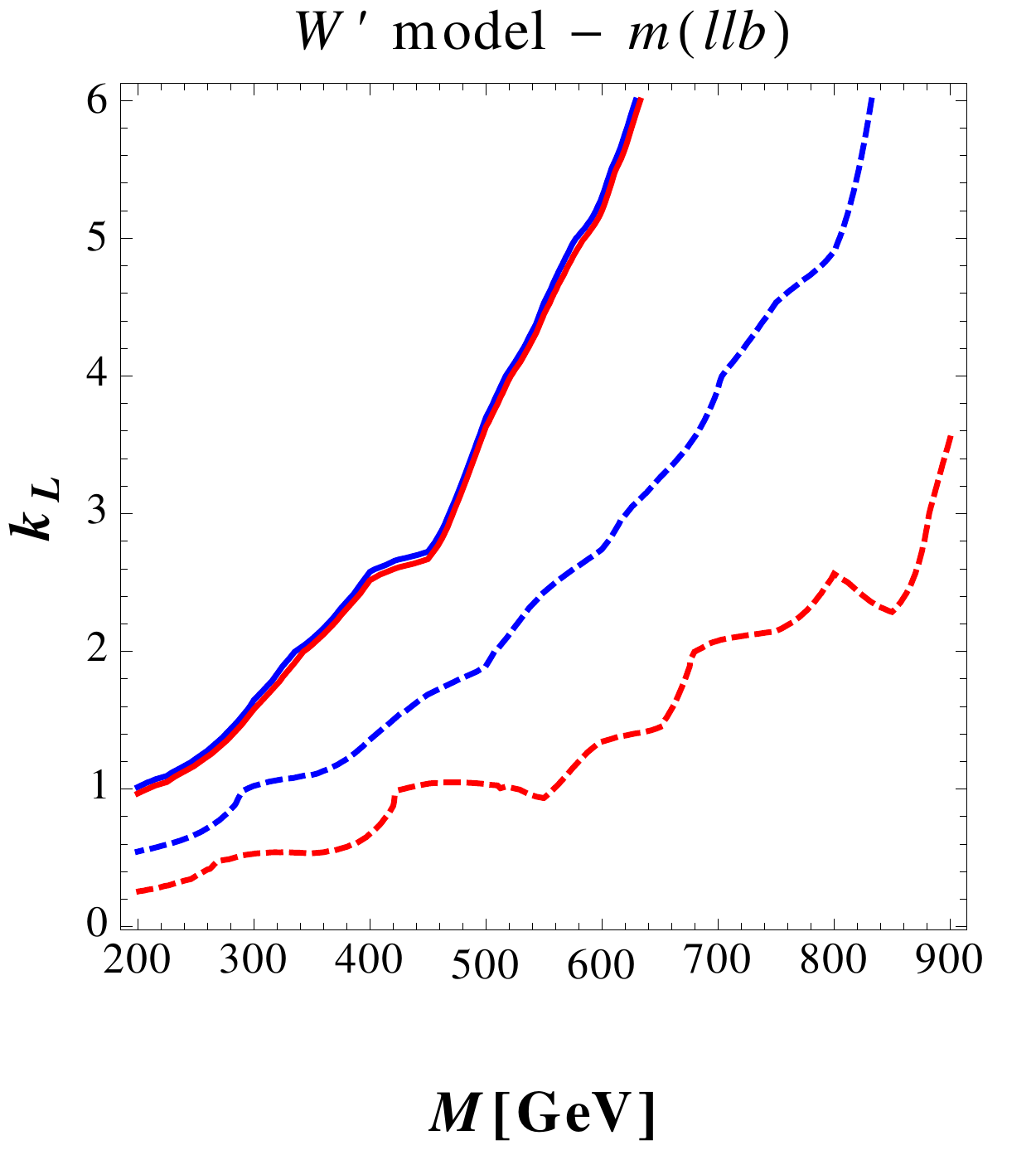} 
\includegraphics[scale=0.62]{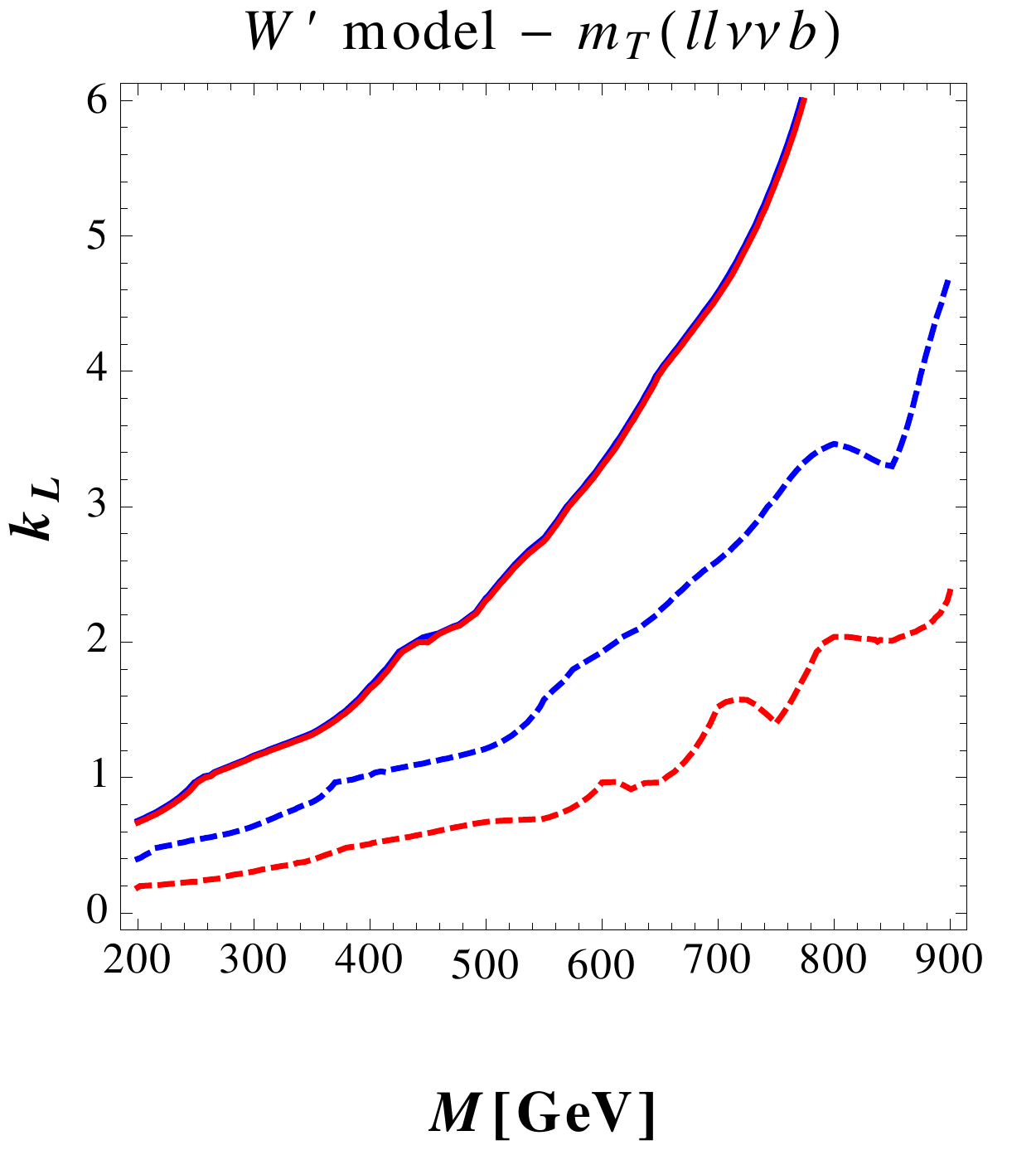} 
\caption{95\% CL expected exclusion regions in the $M$ vs $k_L$ plane: the left plot makes use of the $m(llb)$ distribution while the right plot makes use of the $m_T(ll\nu \nu b)$ distribution. The solid (dashed) blue curves correspond to the pessimistic (optimistic) scenario at 300 fb$^{-1}$ while the solid (dashed) red curves correspond to the pessimistic (optimistic) scenario at 3000 fb$^{-1}$.} \label{bexpwp}
\end{figure}

We see that the best sensitivity is still obtained by considering the $m_T$ differential distribution. Looking at the plots we can see that for both $m(llb)$ and $m_T(ll\nu \nu b)$ can provide significant improvements over current data in the optimistic scenario, but that the pessimistic scenario does not provide any improvement. 

\section{Improving expected bounds with differential ratios}
\label{sec:ratios}
In this section we construct new ratio observables inspired by~\cite{Greljo:2017vvb} and~\cite{Kamenik:2018nxv} and based on the differential distributions we have previously used in our study. In particular we consider the following differential $e/\mu$ ratios
\begin{equation}\label{ratios}
R_x=\left. \frac{d\sigma(pp\to b e^+ \nu_e e^- \bar \nu_e)}{dx} \right/ \frac{d\sigma(pp\to b \mu^+ \nu_\mu \mu^- \bar \nu_\mu)}{dx}
\end{equation}
computed for the observables presented in section IV-A with the same binning as Table~\ref{measdiff}. In the SM, these differential ratios are $R_x^{\rm SM} \simeq 1$. Generic BSM is will affect the first and second generation differently, and below we explore how observables in the class of Eq.~\eqref{ratios} can be used to probe these models. In particular, we use the parameterization of Sections~\ref{sec:EFT} and~\ref{sec:models} where the BSM fields only couple to first generation leptons. These ratio variables are particularly useful because they will have small total uncertainties since many of the systematic ones cancel in the computation of the ratios themselves. We have also checked explicitly that NLO QCD corrections cancel out of the ratio to very high precision.

The SM prediction for $R$ is not exactly one because of QED radiation, namely effects of order $\alpha \log(E/m_\ell)$, where $E$ is the typical energy of the process and $m_\ell$ being the mass of the lepton, which is course very different for the electron and muon. These will be $\mathcal{O}(\text{few \%})$ for $R$ and we use \texttt{PYTHIA8} to do a leading log calculation of these effects. The SM prediction for $R_x$ is given in Table~\ref{ratioSM} where the uncertainties are due to Monte Carlo statistics. We have also explicitly checked that when turning off QED radiation, \texttt{PYTHIA8} predicts $R=1$. 

\begin{table}[h!]
\begin{center}
\vspace{1cm}
\begin{tabular}{|c| c c c c c c |} 
\hline 
$E(llb)$ bin [GeV] & [50,175] & [175,275]& [275,375] & [375,500] & [500,700] & [700,1200]\tabularnewline
$R$ & 0.996 & 0.982 & 0.974 & 0.978 & 0.953 & 0.973 \tabularnewline
Uncertainty & 0.018 & 0.009 & 0.011 & 0.013 & 0.016 & 0.020 \tabularnewline
\hline 
$m_T(ll\nu\nu b)$ bin [GeV] & [50,275] & [275,375]& [375,500] & [500,1000] & & \tabularnewline
$R $ & 0.9823 & 0.967 & 0.955 & 0.991 & & \tabularnewline
Uncertainty  & 0.0064 & 0.010 & 0.018 & 0.027 & & \tabularnewline
\hline 
\end{tabular}
\caption{\label{ratioSM} SM prediction for the flavour ratio in Eq.~\eqref{ratios} taken from 1.2 M Monte Carlo events. Uncertainties are purely statistical.} 
\end{center}
\end{table}

We can now use this variable to get projected limits with a given quantity of LHC data in the mass vs.~coupling plane of our two simplified models. We do this with a $\chi^2$ statistic:
\begin{equation}
\chi^2(g,m) = \sum_i\frac{\left(R_i (\text{SM}) - R_i(g,m)\right)^2}{\sigma_i^2(\text{SM})+\sigma_i^2(g,m)},
\label{eq:chi_ratio}
\end{equation}
where $R_i(g,m)$ is the ratio for a given observable in a given bin, and for a given value of coupling and mass. $R_i (\text{SM})$ is the SM prediction which is given in Table \ref{ratioSM}. The uncertainties are purely statistical, and controlled by the size of the MC sample in our study. Assuming Poisson statistics, the errors are given by
\begin{equation}
\sigma_i^2 = R_i^2\left(\frac{1}{N_e}+\frac{1}{N_\mu}\right)
\end{equation}
where $N_e$ ($N_\mu$) are the number of electron (muon) events in the $i$th bin. This formula assumes that Monte Carlo statistics dominate the uncertainty, which is a good approximation for this variable where most of the higher order corrections cancel or are small. 

\begin{figure}[h!]
\vspace{0.5cm}
\includegraphics[scale=0.6]{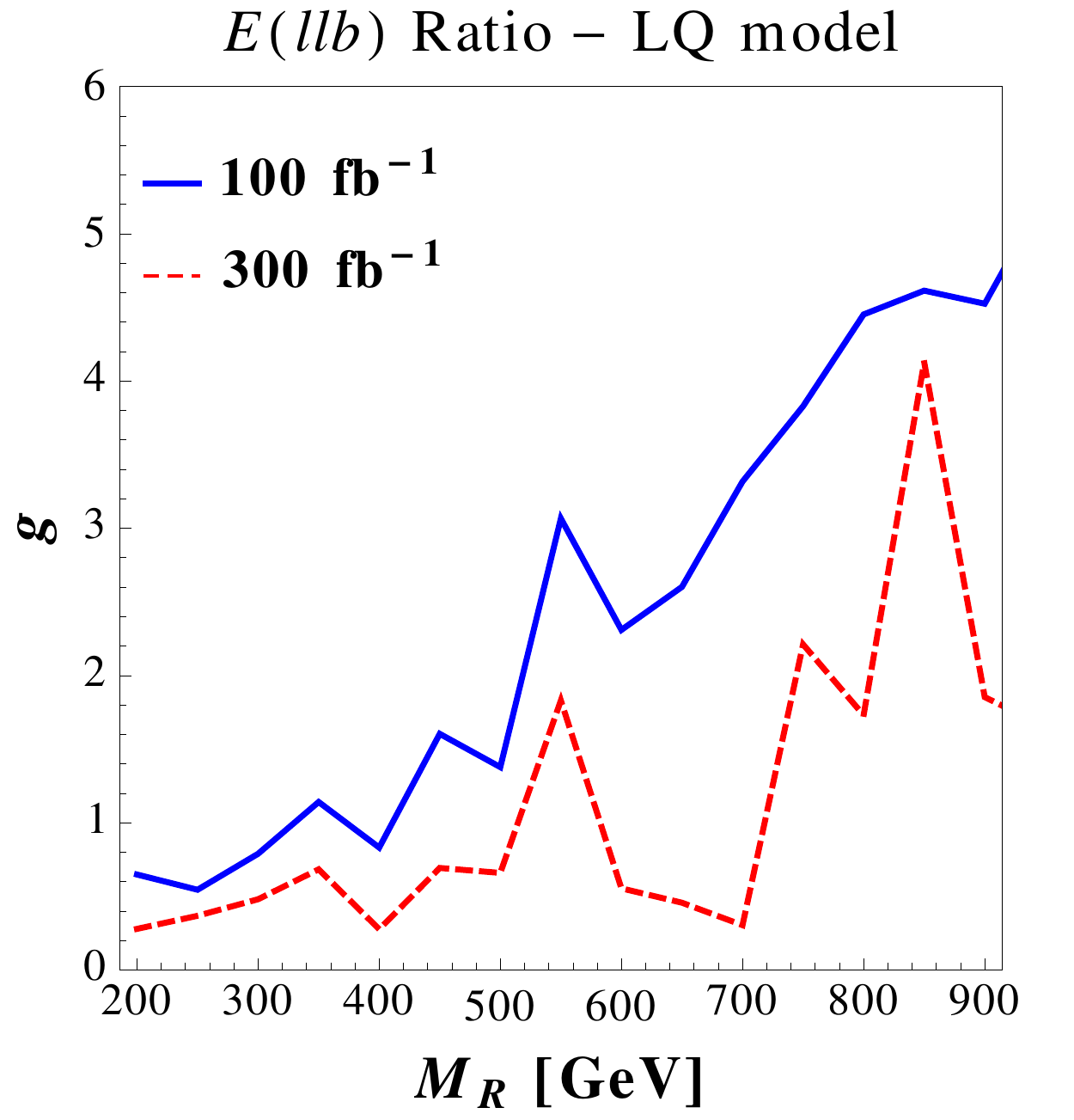} 
\includegraphics[scale=0.6]{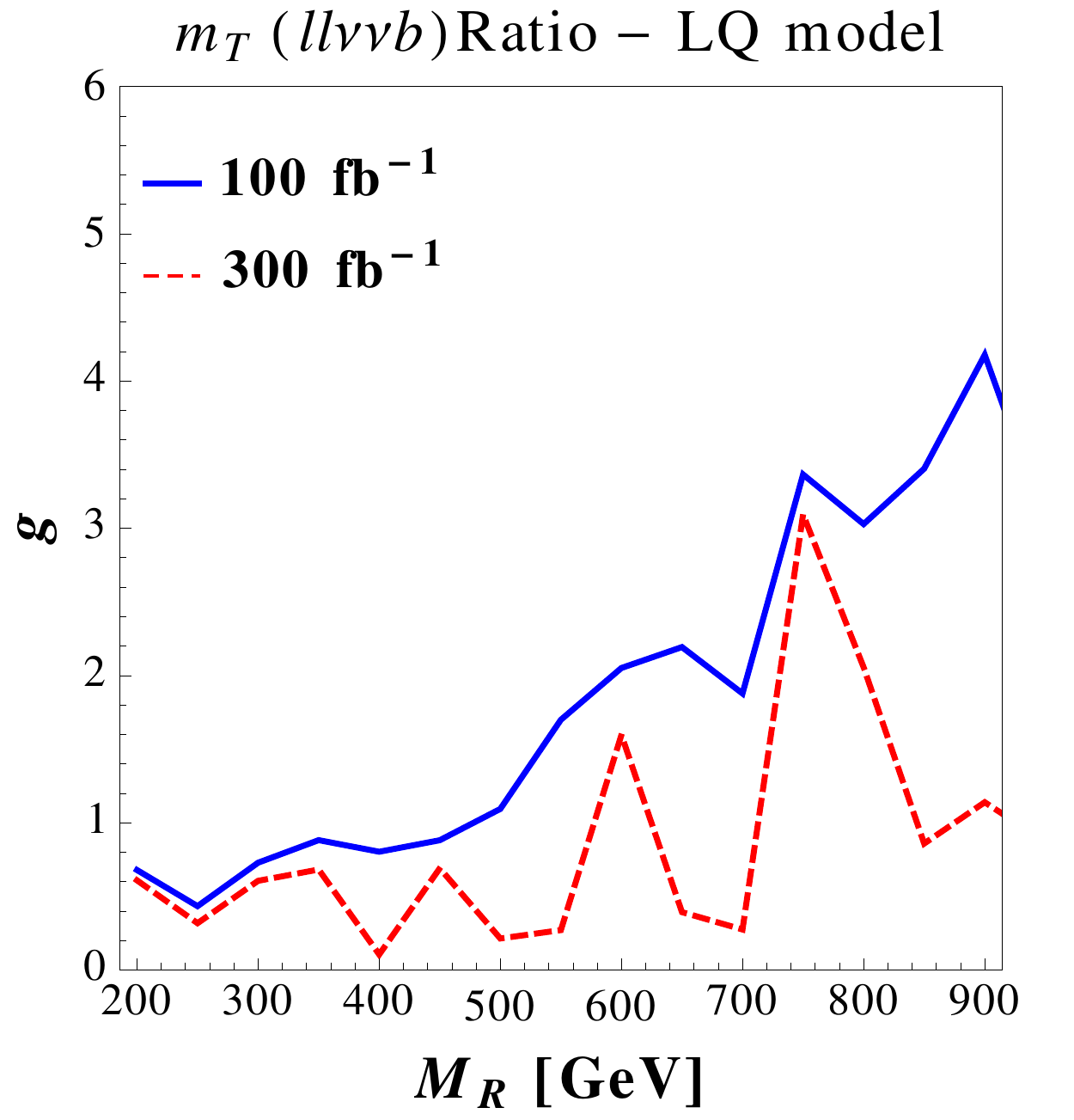} 
\includegraphics[scale=0.6]{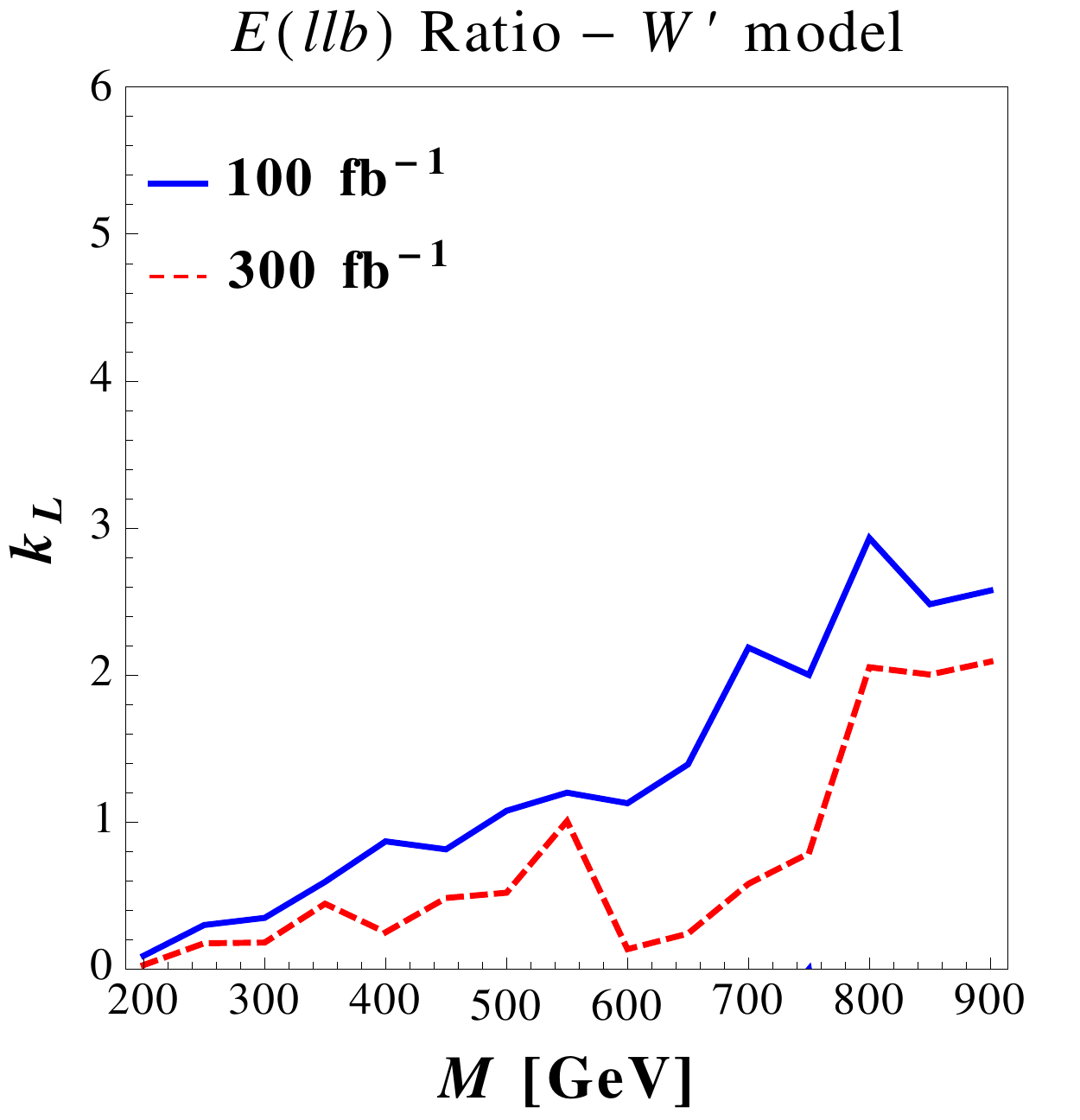} 
\includegraphics[scale=0.6]{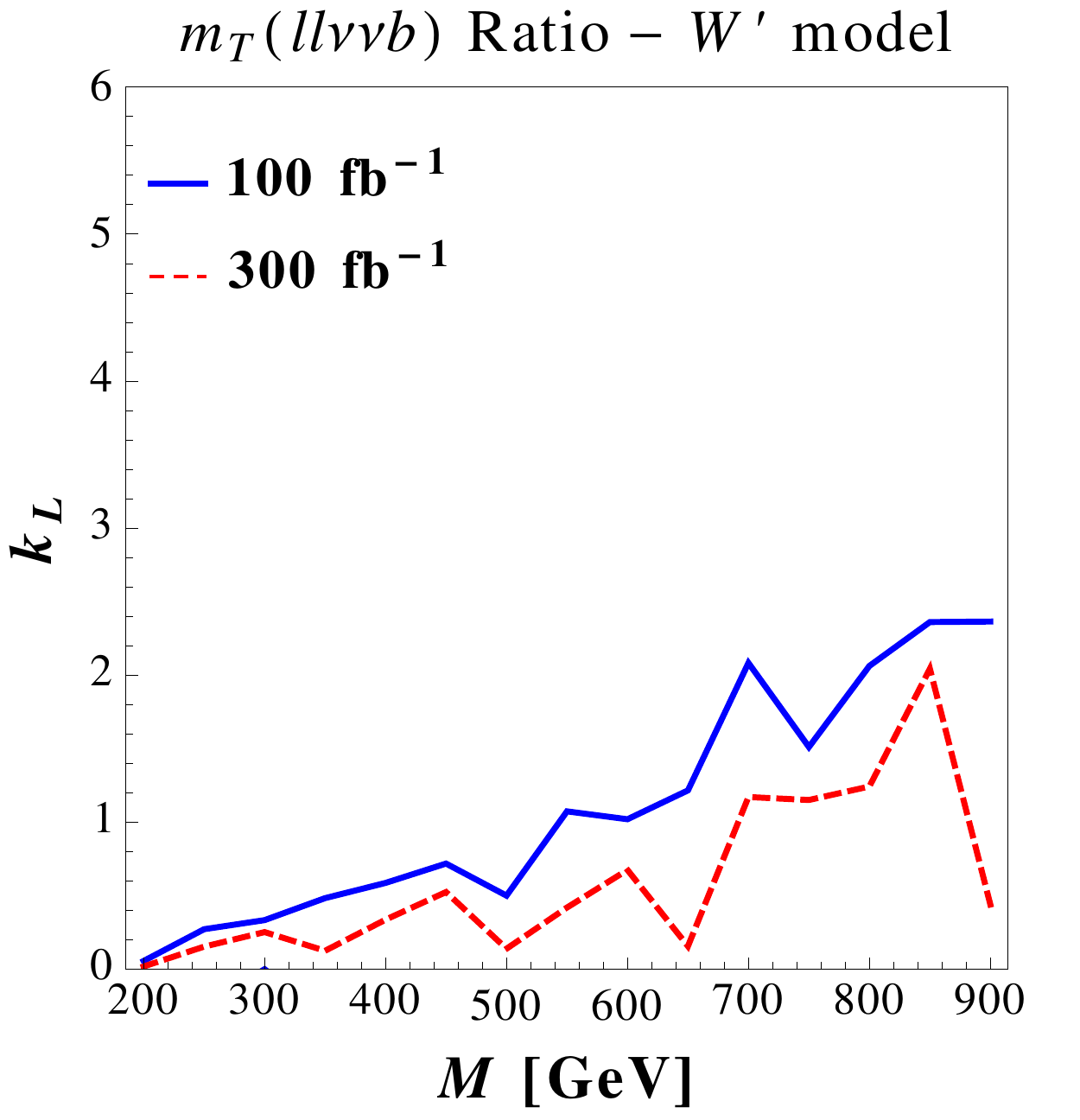} 
\caption{Projected  95\% exclusion limits using the ratio $R_x$ given in Eq.~\eqref{ratios} assuming errors are purely statistical. The left (right) plot is for $x=E(\ell\ell b)$ ($m_T$), and the blue solid lines (red dashed) are for 100 (300) fb$^{-1}$. The top (bottom) row is for the leptoquark ($W'$) model.} \label{fig:ratios}
\end{figure}

We show the projected 95\% exclusions with these ratios in Fig.~\ref{fig:ratios}. We are only able to simulate 100k Monte Carlo events (equivalent to $\sim 100$ fb$^{-1}$) for each point for the BSM scenarios, and those exclusion curves are shown in the solid blue for the leptoquark and $W'$ models. As in Section~\ref{sec:results}, we find that $m_T$ places the strongest constraints, but we also present constraints using $E(\ell\ell b)$ because it does not rely on missing energy and will be complimentary. We find that unlike in Section~\ref{sec:results}, $E(\ell\ell b)$ gives stronger constraints than $m(\ell\ell b)$.

We also present expected limits for 300 fb$^{-1}$ of data in the red dashed. Because of the limitations of our Monte Carlo production, we simply reduce the statistical errors by hand by a factor of $\sqrt{3}$ to estimate the reach with more data, but the central value of our prediction for a given $(g,m)$ BSM parameter point will have larger uncertainties. Therefore, these 300 fb$^{-1}$ curves should be thought of as a crude estimate. With the high luminosity LHC data (3,000 fb$^{-1}$), better limits are expected, but a more precise error budgeting is necessary to make quantitative statements, and we leave this to future work.

\section{Conclusions}
\label{sec:conc}
New physics that couples third generation quarks to first and second generation leptons is as yet very weakly constrained. In this work, we have used the recent ATLAS measurements~\cite{Aaboud:2017qyi}  of differential cross-sections of single top quark production in association with a $W$ boson to constrain this scenario.
We have first parametrized new physics effects by a set of three effective four-fermion operators shown that the $m(llb)$ and $m_T(ll\nu \nu b)$ differential distributions place the strongest current bounds on EFT coefficients coupling the third generation of quarks to leptons. We have parameterized new physics as coupling only to electrons, but more general couplings to electrons and muons can be obtained with simple rescaling.  
We have also computed future expected bounds on this scenario at 300 and 3000 fb$^{-1}$.

The limits found on EFT Wilson coefficients turn out to lie outside the regime of validity of the EFT for perturbative couplings. Therefore, in order to get a strictly valid analysis we have also considered two simplified models, scalar LQ and sequential $W'$, that can be matched onto our EFT operators at low energy. We have derived current and expected bounds in the {\it mass vs coupling} plane  by using the same $m(llb)$ and $m_T(ll\nu \nu b)$ differential distributions. We have compared current bounds with the EFT limits and obtained that, in the $W'$ case, the EFT bounds turned out to be weaker than from the full theory, while the opposite is true in the LQ case. In all these cases, the best sensitivity is obtained by considering the $m_T$ differential distribution.
Our computations of expected limits from future data show that for both $m(llb)$ and $m_T(ll\nu \nu b)$, there can only be very limited improvements on the bounds unless the systematic uncertainties on the measurements can be reduced with more data.
 
In these new physics scenarios, one generically expects different coupling to electrons and muons. If we make the assumptions that new physics couples dominantly to a single lepton and use the very good lepton flavour universality of the Standard Model, ratios of differential distributions of different flavours can be used to place significantly stronger constraints. In these ratios defined in Eq.~\eqref{ratios}, the dominant systematic uncertainties will cancel out to a good approximation, and data can be used as a control sample rather than relying on theory and Monte Carlo predictions. We find that the $m_T(ll\nu \nu b)$ and $E(\ell\ell b)$ differential distributions considered in the ATLAS paper can place much stronger limits on the parameter space than traditional differential distributions, and we have estimated the sensitivity at 100 and 300 fb$^{-1}$. As before, the $m_T$ observable gives the strongest limits, but it also relies on missing energy, so we present both observables as they are complimentary. 

New physics coupling third generation quarks to first and second generation leptons has recently received significant attention because of the persistent $R_K$ and $R_{K*}$ anomalies. The EFT operators in Eq.~\eqref{eftops} can be used to explain these anomalies~\cite{Alonso:2015sja}. Using a simple CKM scaling to to relate the $t-b$ operators here to the $b-s$ operators needed to explain the anomaly, the best fit values of $C_1$ and $C_2$ turn out to be about $3$ orders of magnitude smaller than the sensitivity of the present analysis. On the other hand, its possible that more complicated flavour structures may allow the bounds from the top sector to constrain models that explain these anomalies. 

The BSM scenarios considered here are as yet mostly unconstrained. While the EFT analysis is a useful way to classify BSM scenarios, we find that the current data does not put sufficiently strong constraints to be safely within the regime of validity of the EFT. Regardless, this methods presented here exploiting differential distributions and ratios of differential distributions can probe as yet unexplored regimes of BSM physics.

\section*{Acknowledgments}
We are grateful to Kevin Finelli, Dag Gillberg, and Alison Lister helping us understand the details of~\cite{Aaboud:2017qyi}. 
AT would like to thank Hua-Sheng Shao for the help with MadGraph5\_aMC@NLO and PYTHIA showering.
This work is supported in part by the Natural Sciences and Engineering Research Council of Canada (NSERC).

\appendix

\section{Simulation and fitting procedure}
\label{app:fitting}
In the EFT case, we first set $C_2=0$ and consider 33 unequally separated  points  for $C_1$ in the range $[-20,20]$ TeV$^{-2}$ and generate 100k events for each value of $C_1$. We then set $C_1=0$ and consider 63 unequally separated points for $C_2$ in the range $[-40,40]$ TeV$^{-2}$ and generate 100k events for each value of $C_2$. In both scenarios, after applying the selection cuts described in Section \ref{sec:sim}, we construct $m(llb)$ and $m_T(ll\nu\nu b)$ distribution histograms by rescaling each bin using the $k$-factors of Table \ref{kfactors}. We then use such histograms to compute the chi-squared in Eq.~\eqref{chisq}. For the coefficient $C_1$ we fit the chi-squared values with an even fourth order polynomial, namely $\chi^2(C_1)=a_0+a_2C_1^2 +a_4C_1^4$. This choice is due to the fact that the EFT operators do not interfere with SM. While, for the coefficient $C_2$ we fit the chi-squared values with an generic fourth order polynomial, namely  $\chi^2(C_2)=b_0+b_1C_2+b_2 C_1^2 + b_3C_2^3+b_4C_2^4$. The choice of this function is due to the fact that the EFT operators in this case do interfere with SM and therefore linear and cubic terms are allowed. 

The generated Monte Carlo sample is sufficient to set bounds using current measurements at 36.1 fb$^{-1}$, and the resulting fitting functions for the chi-squared are shown in Fig.~\ref{boundeftcurrent}. Because of the limitations of our Monte Carlo resources, to compute expected bounds at 300 and 3000 fb$^{-1}$ we used the same 100k events.\footnote{To properly capture the statistical uncertainties, one should simulate ${\cal O}(1$M) events for each new physics parameter point.} This approximation is reasonable since the total uncertainty entering in the chi-squared computation is dominated by systematics in both cases. The resulting fitting functions used to establish the expected bounds have been shown in Fig.~\ref{boundeftexpected}.

For the scalar LQ model we consider a non uniform grid of 105 points in the $M_R$ vs. $g$ plane, where $M_R\in[200,900]$ GeV and $g\in [0,8]$. While for the $W'$ model we consider a non uniform grid of 70 points in the $M_{W'}$ vs. $k_L$ plane, where $M_{W'}\in[200,900]$ GeV and $k_L\in [0,6]$. We generated in both models 100k events for each point of the scan and after applying the selection cuts described in Section \ref{sec:sim} we construct $m(llb)$, $m_T(ll\nu\nu b)$ and $E(llb)$ distribution histograms (by rescaling each bin using the $k$-factors of Table \ref{kfactors}). We then use $m(llb)$, $m_T(ll\nu\nu b)$ distributions to compute the chi-squared of Eq.~\eqref{chisq}. We interpolate the chi-squared values with a continuous function of mass and coupling which we use to determine the 95\% CL contours. The LQ 95\% CL contours are shown in Fig.~\ref{boundlq} and \ref{bexplq}, while the $W'$ 95\% CL contours are shown in Fig.~\ref{boundwp} and \ref{bexpwp}.

To compute projected limits for the ratio variables of Sec.~\ref{sec:ratios}, we use a slightly different procedure because the errors are now dominated by statistical uncertainties. Beginning with the same MC events, we find the $\chi^2(g,m)$ statistic from Eq.~\eqref{eq:chi_ratio} for a fixed mass $m$ as a function of coupling $g$, requiring that $\chi^2(0,m)=1$. We then do a linear interpolation between the points and find the value of $g$ that is excluded at 95\%. We then do a linear interpolation for different masses to get the curves in Fig.~\ref{ratioSM}. We have generated $\sim 100$ fb$^{-1}$ for each BSM point, so the statistical errors in Eq.~\eqref{eq:chi_ratio} and central values are consistent. In order to estimate the limit for 300 fb$^{-1}$, we reduce the errors by hand by $1/\sqrt{3}$ and keep the central values the same. These central values are now not guaranteed to be within the error of the true value, so we must take these 300 fb$^{-1}$ to be crude estimates of the projected limits.

\end{document}